\documentclass[twocolumn]{aastex701}
\usepackage{graphicx}
\usepackage{float}
\usepackage[utf8]{inputenc}
\usepackage{natbib}
\usepackage{lmodern}
\usepackage{hyperref}

\submitjournal{ApJ}
\usepackage{CJK}

\shorttitle{WD\,J0234$-$0406}
\shortauthors{BZ}

\begin{document}
\begin{CJK*}{UTF8}{gbsn}
\title{High Velocity Circumstellar Gas Orbiting a White Dwarf Star}

\correspondingauthor{BZ}
\email{ben@astro.ucla.edu}

\author[0000-0001-6809-3045]{B. Zuckerman}
\affiliation{Department of Physics and Astronomy, University of California, Los Angeles, Los Angeles, CA 90095-1562, USA}
\email{ben@astro.ucla.edu}

\author[orcid=0000-0002-3307-1062,sname='Le Bourdais']{\'{E}rika Le Bourdais}
\affiliation{Trottier Institute for Research on Exoplanets and D\'epartment de physique, Universit\'e de Montr\'eal, Ave. Th\'er\`ese-Lavoie-Roux Montr\'eal, Québec H2V 0B3, Canada}
\affiliation{Centre de recherche en astrophysique du Qu\'ebec, Montr\'eal, Qu\'ebec, Canada}
\email{erika.le.bourdais@umontreal.ca}

\author[0000-0001-5854-675X]{Beth L. Klein}
\affiliation{Department of Physics and Astronomy, University of California, Los Angeles, Los Angeles, CA 90095-1562, USA}
\email{kleinb@ucla.edu}

\author[0000-0003-4609-4500]{Patrick Dufour}
\affiliation{Trottier Institute for Research on Exoplanets and D\'epartment de physique, Universit\'e de Montr\'eal, Ave. Th\'er\`ese-Lavoie-Roux Montr\'eal, Québec H2V 0B3, Canada}
\affiliation{Centre de recherche en astrophysique du Qu\'ebec, Montr\'eal, Qu\'ebec, Canada}
\email{patrick.dufour@umontreal.ca}

\author[0000-0001-9834-7579]{Carl Melis}
\affiliation{Department of Astronomy and Astrophysics, University of California, San Diego, La Jolla, CA 92093-0424, USA}
\email{cmelis@ucsd.edu}

\author[0000-0001-6654-7859]{Alycia J. Weinberger}
\affiliation{Earth and Planets Laboratory, Carnegie Institution for Science, 5241 Broad Branch Rd. NW, Washington, DC 20015, USA}
\email{aweinberger@carnegiescience.edu} 

\author[orcid=0000-0002-8808-4282]{Siyi Xu (许\CJKfamily{bsmi}偲\CJKfamily{gbsn}艺)} 
\affiliation{Gemini Observatory/NOIRLab, 950 N Cherry Ave., Tucson, AZ 85719}
\email{siyi.xu@noirlab.edu}

\author[0000-0002-2384-1326]{Antoine B{\'e}dard} \affiliation{Department of Physics, University of Warwick, Coventry, CV4 7AL, UK }
\email{antoine.bedard@warwick.ac.uk}

\author[0000-0002-6164-6978]{Detlev Koester}
\affiliation{Institut f\"{u}r Theoretische Physik und Astrophysik, University of Keil, D-24098 Kiel, Germany}
\email{koester@astrophysik.uni-kiel.de}

\begin{abstract}

Numerous white dwarf stars are known to be orbited by disks of gas and dust.To date, broad, $\sim$300 km s$^{-1}$ wide, gaseous circumstellar absorption features have only been reported for the already iconic WD\,1145+017, where one is witnessing the breakup of an extrasolar asteroid in real time. We report here the discovery of absorption from circumstellar gas around a second white dwarf (WD\,J0234–0406) with similarly broad features. The observed lines are carried by ions of Ca, Cr, Fe, Ti, Mg, Mn, Na, O, Si, Sc, Sr, Ti, and V. In addition, deep, non-photospheric lines of \ion{Si}{4} are seen in the ultraviolet; we compare these with \ion{Si}{4} lines previously seen in the ultraviolet spectra of various other white dwarfs. The apparent broadband flux of WD\,1145+017 is known to change often and rapidly as chunks of the asteroid pass between the star and Earth. No such variations are seen in the brightness of WD\,J0234–0406. In addition, while the strength/structure of circumstellar absorption features at WD\,1145+017 has changed dramatically with time, nothing similar is seen at WD\,J0234–0406. Excess infrared emission at WD\,J0234-0406 indicates the presence of circumstellar dust particles. 
\end{abstract}


\section{Introduction}\label{sec:intro}
\end{CJK*}
Optical and ultraviolet spectroscopy have revealed the presence of elements heavier than helium in the photospheres of 25$\%$ - 50$\%$ of white dwarf stars \citep{zuckerman_metal_2003, zuckerman_ancient_2010, koester_frequency_2014}. A much smaller percentage shows evidence for orbiting dust particles, and an even smaller percentage for orbiting gas. These various phenomena are now understood to be, typically, the consequences of the accretion of rocky minor planets (asteroids) by the stars \citep{veras_post-main-sequence_2016, veras_planetary_2021,farihi_circumstellar_2016, jura_extrasolar_2014,deeg_characterizing_2018}. Sometimes the accreted body could be a Kuiper-Belt analog \citep{xu_chemical_2017}, a large moon \citep{doyle_icy_2021}, or even a major planet \citep{jura_x-ray_2009}. 

Direct evidence for orbiting gas usually takes the form of emission lines from one or a variety of elements \citep[e.g.,][]{gansicke_gaseous_2006,melis_serendipitous_2020,dennihy_five_2020, gentile_fusillo_white_2021}. These lines often display a wide range of radial velocities indicative of rapid orbital motions in a disk. Circumstellar absorption features are much rarer, in part because the orbiting material must lie in a plane that intersects the line of sight from Earth to star. For nearly circular circumstellar orbits, one would anticipate a narrow absorption line. To date, examples of a few such narrow lines have been presented in \citet{debes_detection_2012}, \citet{steele_characterization_2021}, \citet{gansicke_chemical_2012} and \citet{vanderbosch_recurring_2021}. When there is evidence for circumstellar dust or gas at a white dwarf, there are also absorption lines from one or more elements in the photospheric spectrum of the star. The interpretation is that the dust and gas accrete onto the star and remains in the photosphere for a while in the form of various ions before it sinks into the interior out of sight.

White dwarf WD\,1145+017 differs from those stars mentioned in the preceding paragraph in that it displays numerous absorption features, each of which covers a wide range of velocities \citep[$\sim$300 km s$^{-1}$,][]{xu_evidence_2016}. This indicates that the orbits of the absorbing gas must have a substantial radial component. Models for how this might occur are considered in \citet{budaj_wd_2022} and references therein. Substantial variations in the broadband flux of WD\,1145+017 were first discovered in the analysis of Kepler K2 data by \citet{vanderburg_disintegrating_2015}. These variations have been interpreted as due to the passage between the star and Earth of chunks of a disintegrating asteroid. The gaseous absorption features' strengths and radial velocities at WD\,1145+017 also vary with time \citep{redfield_spectroscopic_2017,cauley_evidence_2018,xu_shallow_2019,fortin-archambault_modeling_2020,le_bourdais_revisiting_2024}.

In this paper, we consider a second white dwarf that displays numerous broad circumstellar absorption lines from a variety of elements. The Gaia DR3 position of the white dwarf is 02 34 15.51; -04 06 09.28. The only published analysis of the spectrum of the star is by \citet[hereafter GF21]{gentile_fusillo_white_2021}, who denote it as WD\,J0234-0406. Here, we use the same name although the star has been previously identified in the literature as PHL\,1360.

GF21 found that WD\,J0234-0406 has an unusually large H mass fraction compared to other He-rich white dwarfs and they measured extremely wide \ion{Ca}{2} IR triplet profiles in emission. Quoting from their paper, ``the emission profiles appear relatively symmetric that could suggest a circular disc.'' They do not report any non-photospheric absorption lines. So the principal addition of the present paper is the discovery of the broad absorption features; these indicate that the orbiting disk is far from circular.
 
As discussed below, WD\,J0234-0406 and WD\,1145+017 have some common features, but also some that differ. Perhaps the most striking difference is the lack of measurable secular variations in both the broadband flux and the circumstellar line characteristics from WD\,J0234-0406 (see Section~\ref{sec4p2}). This suggests that the asteroid(s) that were the parent body for the observed gas and dust have already been broken into tiny pieces, well distributed in the circumstellar material. 

Two unusual absorption lines seen in the ultraviolet spectrum of WD\,J0234-0406 arise from \ion{Si}{4}. Because the photosphere of the star is too cool to support this ionization state, the observed gas must be associated with the circumstellar material. For comparison with WD\,J0234-0406, we gather together some spectral characteristics of the handful of other white dwarfs that display these lines. 

Section~\ref{sec2} presents a discussion of our observations and data reduction. Section~\ref{sec3} summarizes results for circumstellar elements and for those in the photosphere. We discuss these results in Section~\ref{sec4}. Section~\ref{sec5} presents the consideration of a few interesting aspects of the spectrum of WD\,J0234-0406, including the ultraviolet lines of \ion{Si}{4}. Section~\ref{sec6} gives our conclusions.

\section{OBSERVATIONS AND DATA REDUCTION}\label{sec2}

Our 2017 discovery of the noteworthy spectrum of WD\,J0234-0406 was one product of our team's large scale survey to identify heavily accreting white dwarf stars \citep{melis_first_2018,doyle_new_2023}. As may be seen in Table~\ref{tab1:obs}, follow up observations with a variety of spectrometers were carried out in 2018, and then, after a break of a year, WD\,J0234-0406 was extensively observed with HIRES. The GF21 observation of WD\,J0234-0406 occurred in 2019 January.

\begin{deluxetable*}{lccccc}
\tablecaption{Observation Log \label{tab1:obs}}
\tablehead{\colhead{UT Date} & \colhead{Instrument} & \colhead{Wavelength Range} & \colhead{Resolving Power} & \colhead{Exposure} & \colhead{S/N}\\[-4pt]
\colhead{} & \colhead{} & \colhead{(\AA)} &  & \colhead{(s)} & \colhead{}
}
\startdata
2017 Dec 9 & Kast blue & 3450--5475 & 1300  & 3600   & 38 \\
2017 Dec 9 & Kast red & 5600--7850 & 3000    & 3600  & 47 \\
2018 Jan 1 & HIRESb & 3113--5944 & 37000 & 3300    & 29 \\
2018 Jul 12  &   COS  &  1135--1425   & 15000  & 19570  &   8   \\
2018 Sep 2  &   COS  &  1135--1425   & 15000  & 19570  &   8   \\
2018 Oct 15 & MagE & 3200--8200 & 7500 & $2 \times 3000$   & 120 \\
2018 Oct 27 & HIRESb & 3113--5944 & 37000 & 3300    & 14 \\
2020 Sep 13 & HIRESb & 3113--5944 & 37000 & $2 \times 1800$ & 36 \\
2020 Oct 7 & HIRESr & 4720--8990 & 37500 & $2 \times 3500$ & 75 \\
2020 Oct 8 & HIRESb & 3113--5944 & 37000 & 3600    & 33 \\
2021 Oct 9 & HIRESb & 3113--5944 & 37000 & $2 \times 3600$ & 50 
\enddata
\tablecomments{Signal-to-noise ratio (S/N) per pixel measured at 1350\,{\AA} in COS spectra, at 3850\,{\AA} in Kast blue spectra and 5800\,{\AA} in red spectra, at 5100\,{\AA} in MagE spectra, and at 5100\,{\AA} in HIRES spectra. COS spectra displayed in the paper have been smoothed over 5 pixels.}
\end{deluxetable*}
\subsection{Kast Spectroscopy}\label{sec2p1}

Discovery observations were performed at Lick Observatory with the Kast Double Spectrograph mounted on the Shane 3\,m telescope. Kast observations simultaneously employed the blue and red arms with light split by the d57 dichroic around 5700\,\AA. After splitting, blue light was passed through the 600/4310 grism, while red light was passed through the 830/8460 grating. A slit width of 1.0$'$$'$ was used.

Kast data are reduced using standard {\sf IRAF} long-slit tasks, including bias subtraction, flat-fielding, wavelength calibration with arc lamps, and instrumental response calibration via observations of flux calibration standard stars. Arc lamp frames were not obtained close in time to science frames and as such, the zero-point of the wavelength scale is not accurate. Wavelength ranges covered, resolving powers obtained, and final spectral signal-to-noise ratios are reported in Table~\ref{tab1:obs}.

\subsection{HIRES Spectroscopy}\label{sec2p2}

 WD\,J0234-0406 was observed with spectrometers on four telescopes (see Table~\ref{tab1:obs}). Most of the data were obtained with the HIRES Echelle Spectrometer \citep{vogt_hires_1994} on the Keck I telescope at Maunakea Observatory on the UT dates given in Table~\ref{tab1:obs}. We used the C5 decker with a slit width of 1.148''. All optical spectra and wavelengths quoted in this paper are in air, corrected to the heliocentric reference frame

Data reduction was performed by using both MAKEE and IRAF following \citet{klein_chemical_2010} and \citet{xu_elemental_2014}. Typically, we used the spectrum of Feige 34 and similar white dwarfs to calibrate the spectrum of WD\,J0234-0406.

\subsection{MagE Spectroscopy}\label{sec2p3}

Observations at Las Campanas Observatory with the MagE spectrometer \citep{marshall_mage_2008} on the Magellan/Baade telescope were obtained for WD\,J0234-0406. Table~\ref{tab1:obs} provides important observational and data quality parameters. The spectrograph was used with a 0$'$$'$.5 slit, and data were reduced with the facility Carnegie Python pipeline \citep{kelson_optimal_2003}. With a combination of IRAF and in-house IDL routines we perform relative flux calibration on individual echelle orders with standard star spectra, then merge orders with the application of small flux offsets when needed to bring overlapping order segments into agreement.

\begin{figure*}
\includegraphics[trim={1cm 20.5cm 1cm 3cm},clip,width=\linewidth]{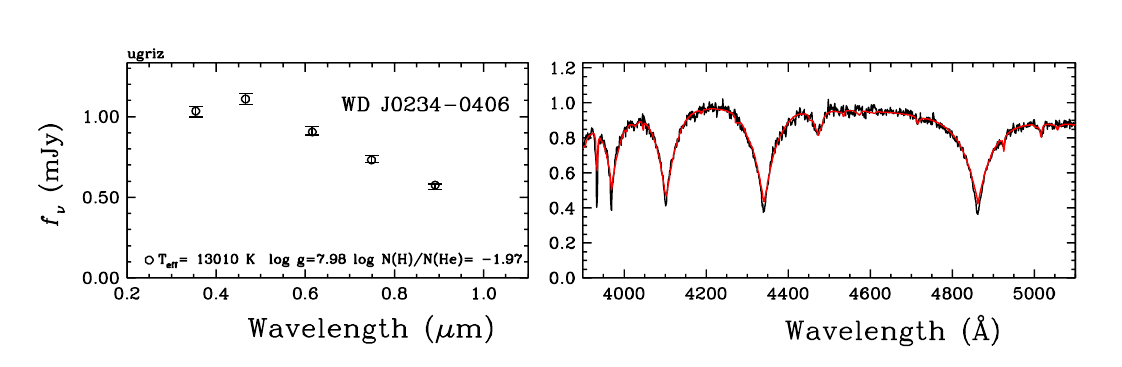}
\caption{Photometric/spectroscopic fit to WD\,J0234-0406. Left panel: photometric fit (open circles) on SDSS \textit{ugriz} data (error bars). Right panel : Best fit model (red) over the Kast spectrum.}
\label{fig1:fit_photo}
\end{figure*}

\subsection{Cosmic Origins Spectrograph Spectroscopy}\label{sec2p4}

$Hubble$ $Space$ $Telescope$ Cosmic Origins Spectrograph \citep[hereafter COS,][]{green_cosmic_2012} observations were obtained for WD\,J0234-0406 during two visits. Seven spacecraft orbits total were allocated to this target as part of GO program 15461. Observations were performed with the G130M grating centered at a wavelength of 1291\,\AA\ and were obtained in the {\sf TIME-TAG} mode using the 2.5$''$ diameter primary science aperture. Two different {\sf FP-POS} values were utilized during the observations to minimize fixed pattern noise in the detector while minimizing the impact of Ly-$\alpha$ geocoronal emission on the detector. This setup yields the spectral resolving power and coverage given in Table~\ref{tab1:obs} and has a gap in spectral coverage from 1274 to 1287\,\AA.

The raw COS data were processed with the {\sf\verb+CALCOS+} pipeline and coadded with the use of the IDL script {\sf \verb+COADD_X1D+} \citep{danforth_hubblecos_2010,keeney_significance_2012}. The final spectrum is flux calibrated and presented in vacuum wavelengths corrected to the heliocentric reference frame. 

The final spectral signal-to-noise ratio is reported in Table~\ref{tab1:obs}. Following \citet{jura_two_2012}, we use the {\sf time filter} module to extract the night-time portion of the data around the \ion{O}{1} lines between 1300 and 1308\,\AA\ to mitigate terrestrial day airglow emission. Lower signal-to-noise spectra are obtained for the \ion{O}{1} lines between 1300 and 1308\,\AA\ as a result of the filtering.

Ultraviolet spectra of WD\,J0234-0406 reveal a highly broadened and deep Ly$\alpha$ photospheric absorption line. Because of the Ly$\alpha$ line, there is essentially no signal below 1300 \AA. This is reminiscent of the significantly broader-than-expected Ly$\alpha$ line found by \citet{xu_chemical_2017} in white dwarf WD\,1425+540. While new models have since been developed that better reproduce ultraviolet hydrogen absorption lines in helium-dominated white dwarf atmospheres \citep{gansicke_broadening_2018}, such models were not in use when the observations of WD\,J0234-0406 were planned. The result is that the observed flux is dramatically lower than originally expected, and a much lower signal-to-noise ratio is obtained than was suggested by preliminary observational planning.

\subsection{Zwicky Transient Facility Photometry}\label{sec2p5}

We queried the ZTF camera at Palomar Observatory \citep{bellm_zwicky_2019,masci_zwicky_2019} for data release 23 to check for photometric variability. Photometry was acquired between July 2018 and December 2024 with the usual seasonal and weather breaks. There were 388, 422 and 71 observations in total in the g$'$, r$'$ and i$'$ bands, respectively. The median ZTF r$'$ and g$'$ cadence during an observing season was one visit every 2 days.

\begin{figure}
\centering
\includegraphics[width=0.99\linewidth,trim={0.1in 0.1in -0.3in -0.2in},clip]{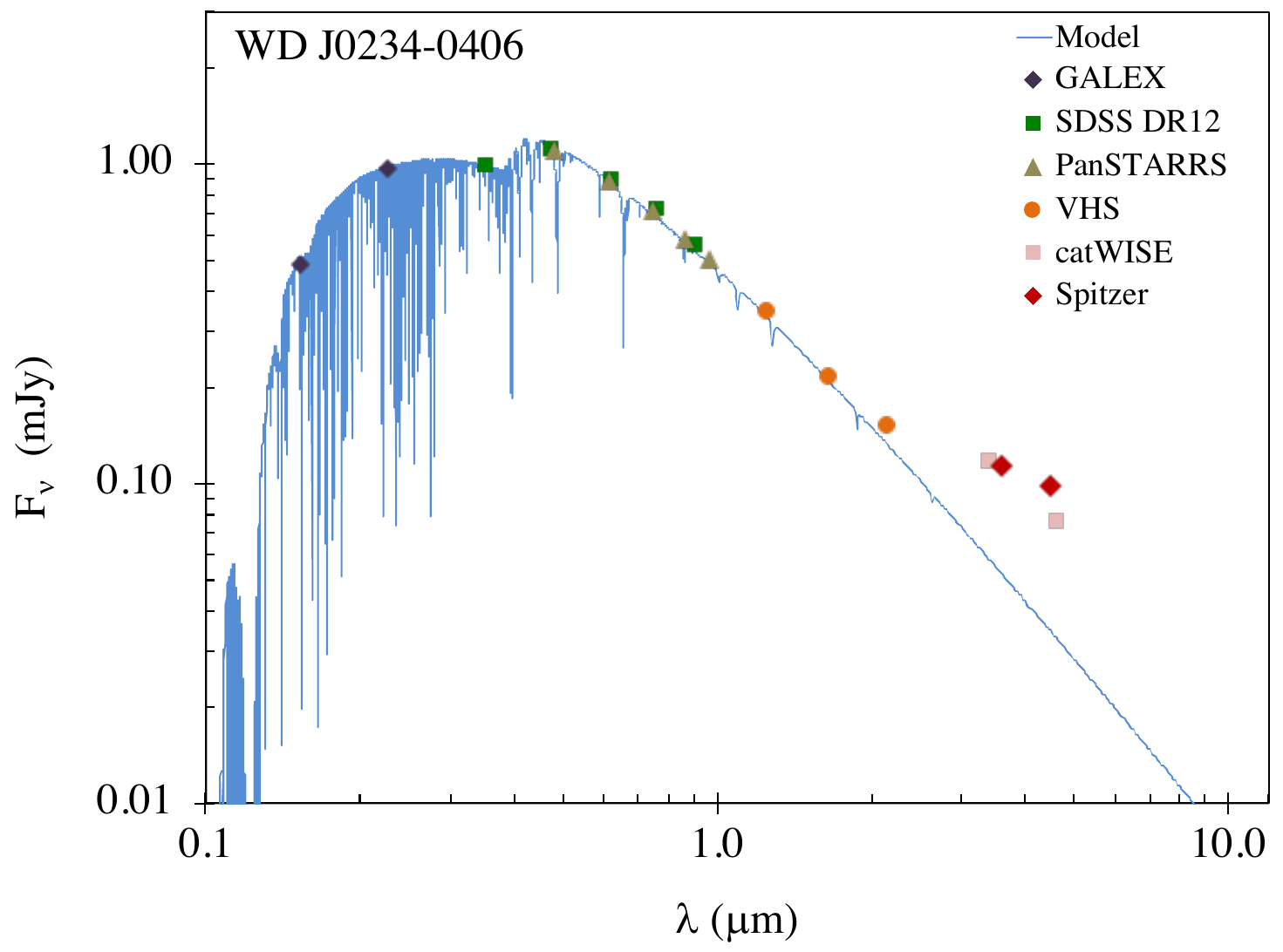}
\caption{Spectral energy distribution of WD\,J0234-0406. 
Photometric data from GALEX \citep{bianchi_revised_2017}, SDSS \citep{alam_eleventh_2015}, Pan-STARRS \citep{flewelling_pan-starrs1_2020}, Two Micron All Sky Survey \citep[2AMSS,][]{cutri_2mass_2003}, VISTA Hemisphere Survey (\citep{mcmahon_first_2013}, and CatWISE \citep{eisenhardt_catwise_2020} surveys. Spitzer data are from \citet{lai_infrared_2021}.}
\label{fig2:SDSSJ0234_SED}
\end{figure}

\section{METHODS AND MODELS}\label{sec3}
\subsection{Photometric Parameters}\label{sec3p1}
The spectral energy distribution for WD\,J0234-0406 is shown in Figures~\ref{fig1:fit_photo} and \ref{fig2:SDSSJ0234_SED}. Excess emission indicative of an orbiting dust disk is evident in Figure~\ref{fig2:SDSSJ0234_SED} in the infrared beginning at a wavelength of about 2~$\mu$m, as first reported in \citet{rebassa-mansergas_infrared-excess_2019}. Minor reddening corrections of GALEX, Sloan Digital Sky Survey (SDSS) and Pan-STARRS photometry were applied with an assumed extinction $E(B-V)=0.01$ for a distance of 92~pc. SDSS to AB corrections were made as prescribed by \citet{eisenstein_catalog_2006}. Uncertainties in the photometry are close to the size of the data points.

Our best-fit stellar model, $T_{\rm eff} = 13,000 \pm 700~{\rm K}$, $\log g = 7.98\pm0.03$, and $\log ({\rm H/He}) = -1.97\pm0.2$, is presented in Figure~\ref{fig1:fit_photo}. The complete physical parameters derived for this white dwarf are listed in Table \ref{tab2:param}. The GF21 model $T_{\rm eff}$ and $\log g$ for the white dwarf are $13,454\pm200$ and $8.01\pm0.03$. Their model H/He ratio agrees with ours. Figure~1 of GF21 displays their spectral energy distribution, to which they fit a model dust disk of blackbody temperature 1226~K, with temperature 1872~K at the disk inner edge.

Although the optical spectrum of WD\,J0234-0406 is dominated by strong Balmer lines, the presence of weak \ion{He}{1} lines indicates that the atmosphere of this white dwarf is actually helium-rich, but with a substantial amount of hydrogen -- similar to the situation in GD 362 \citep{zuckerman_chemical_2007}. Using the model atmosphere code described in detail in \citet{coutu_analysis_2019}, we first computed a grid with $\log g$ between 7.5 and 9.0 in steps of 0.5, $T_{\mathrm{eff}}$ between 11,000 and 16,000~K in steps of 1000~K, and $\log (\mathrm{H}/\mathrm{He}) = -1$ to $-3$ in steps of 0.5. 

Since the low-resolution optical spectrum of WD\,J0234-0406 also shows traces of \ion{Ca}{2} H and K lines, heavy elements were included with an initially estimated calcium abundance, chosen based on prior experience and close to the final value (though the exact starting point is not critical).
All other elements were scaled to calcium following chondritic proportions. The entire grid was recalculated iteratively until a good match to the photospheric metal lines was achieved; this process included approximately accounting for the contribution of the circumstellar disk during the fitting process \citep[see below, and][]{le_bourdais_revisiting_2024}. 

We obtained the atmospheric parameters of WD\,J0234-0406 by simultaneously fitting the SDSS \textit{ugriz} photometry, trigonometric parallax, and the optical spectrum, using a technique similar to that described in detail in \citet{coutu_analysis_2019}. Because the H abundance in WD\,J0234-0406 is unusually large for a He atmosphere white dwarf, obtaining a final solution for this abundance necessitated use of photometry, He 4471 \AA, and Balmer lines. Interestingly, since the depth of the \ion{He}{1} 4471 line is strongly dependent on the H/He ratio, the deduced H/He ratio is driven mostly by a fit to this spectral region.

\begin{deluxetable}{lc}[h]
  \tablecaption{Physical parameters of WD\,J0234-0406.\label{tab2:param}}
  \tablehead{\colhead{Parameter} \hspace{3.5cm} & \colhead{Value}}
  \startdata
  $T_{\rm eff}$ (K)  &  13,010 $\pm$ 710  \\
  $\log g$    &   7.98 $\pm$ 0.01 \\
  Mass ($M_\odot$)    &  0.574 $\pm$ 0.008  \\
  $\log$ H/He   &  -1.97 $\pm$ 0.2  \\
  Gravitational redshift (km s$^{-1}$)   &  28.6 $\pm$ 0.4 \\
  \enddata
\end{deluxetable}

\subsection{Gas Disk Model and Photospheric Abundances}\label{sec:disk-model}\label{sec3p2}

We make use of the updated gas disk model presented in \citet{le_bourdais_revisiting_2024} to reproduce the circumstellar features in both the optical and ultraviolet spectra. The disk model is made of 14 confocal eccentric rings evenly spaced radially with each ring having its own density and temperature. An opacity is calculated in each point of the $50\times50\times14$ 3D grid using \textsc{tlusty} and \textsc{synspec} \citep{hubeny_tlusty_2021}. A Doppler velocity shift is then applied to the opacity to account for the gas rotation and the 3D grid is integrated in the line of sight to obtain 1D opacities as a function of wavelength. The full technical details of the disk model are described at length in \citet{cauley_evidence_2018}, \citet{fortin-archambault_modeling_2020} and \citet{le_bourdais_revisiting_2024}. 

We adjusted the disk's configuration to reflect the shorter extent of radial velocities in the circumstellar absorption features $\pm 50-100 {~\rm km~s^{-1}}$ compared to WD\,1145+017 ($\pm 250-300 {~\rm km~s^{-1}}$). The configuration was based on the \ion{Ti}{2} 3261, 3349 and 3373 {\AA} lines as Ti was the dominant ion in the disk, and the lines were well isolated and uncontaminated by circumstellar absorption lines from other ions. For the disk's temperature and density, we explored a parameter space ranging from 2500 to 7000 K and from $1 \times 10^{-14}$ to $1 \times 10^{-11} {~\rm g/cm}^3$. We find that the solution most suitable for most of the circumstellar features is a midplane disk temperature of 6000 K and a density of $9 \times 10^{-14} {~\rm g~cm}^{-3}$. For the disk's abundances, we started with a CI Chondrite abundance, taking into account the first 30 elements of the periodic table in the model. We then overplotted the combined disk + photosphere model over the spectrum to see how well it matched the circumstellar features. We then adjusted the abundances of the main elements responsible for poorly reproduced features by either increasing the abundance when the feature was too shallow or decreasing when the feature was too deep. The process was done iteratively for all circumstellar features until we obtained an overall satisfying reproduction of all circumstellar features at all wavelengths.

Once the proper disk configuration was obtained, we fitted the photospheric abundances using the method described in \citet{le_bourdais_revisiting_2024}.

\section{RESULTS}\label{sec4}

\subsection{Secular Variability}\label{sec4p2}
\begin{figure}[h]
\centering
\includegraphics[width=1.01\linewidth,trim={0.7cm 6.8cm 1.5cm 8.0cm},clip]{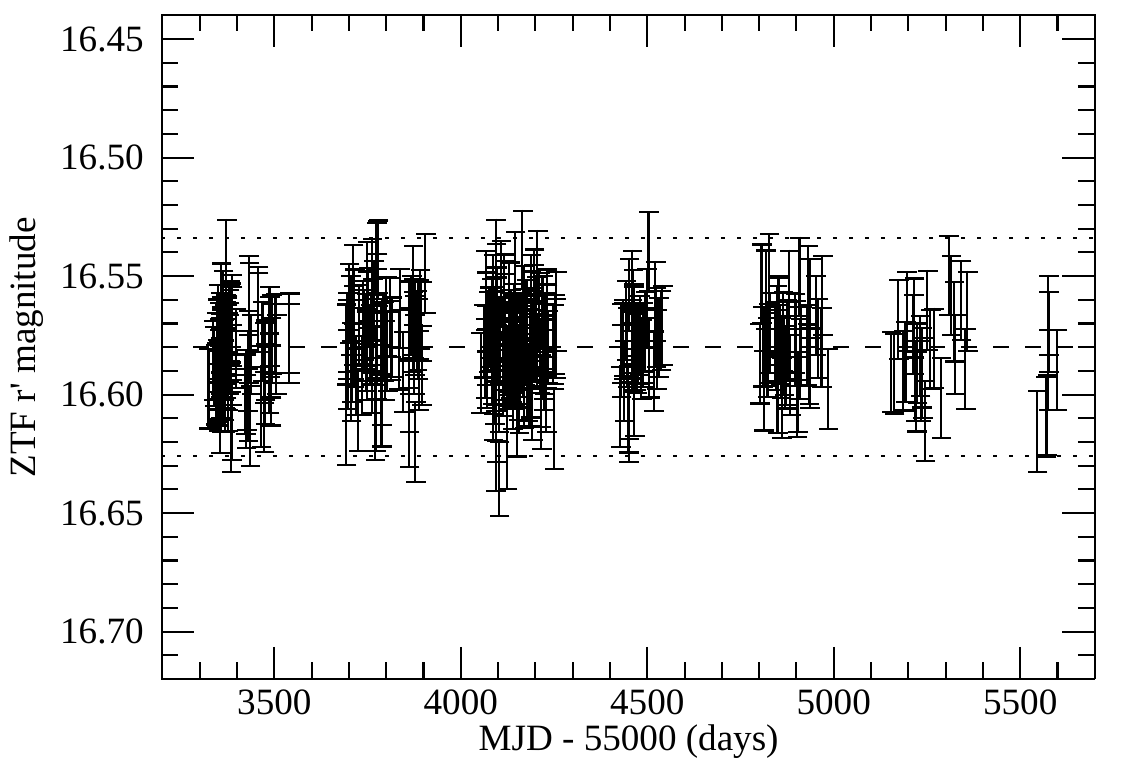}
\caption{ZTF r$'$ lightcurve for WD\,J0234-0406. The star appears to be stable with no obvious transit events or variability. Most spectroscopic observations reported in this paper were performed throughout the first half of the displayed lightcurve time range. \label{fig3:ztf}}
\end{figure}

As noted in the Introduction, there has been no measurable variability in the broadband flux from WD\,J0234-0406. ZTF data were downloaded from the IRSA ZTF DR23 catalog service. Lightcurve data within 5 arcseconds of the stellar position were retrieved. Data points with catalog flags $>$ 0 are removed, as are all observations conducted at an airmass $\geq$1.8. The remaining data points were analyzed. The ZTF photometry yielded g$'$, r$'$ and i$'$-band magnitudes of 16.344$\pm$0.016, 16.579$\pm$0.015 and 16.827$\pm$0.020, respectively. The r$'$ ZTF observations are shown in Figure~\ref{fig3:ztf}. For the sake of completeness, we also examined the Transiting Exoplanet Survey Satellite (TESS) 2 minute cadence photometry of WD\'J0234-0406. TESS observations in Sector 31 (start date of 2020-10-22) covered 25.4 days total with an $\approx$2.8 day gap in the middle of the sector. Data were obtained directly from MAST as produced by the Science Processing Operations Center. The TESS observations reveal a flat lightcurve over the monitoring period with raw rms scatter of $\sim$25\%. Binning the lightcurve produces inter-datapoint scatter closer to what was obtained with ZTF but still reveals no obvious variability.

No variability is apparent in the circumstellar absorption features. We compared various lines of \ion{Fe}{2} and \ion{Ti}{2} in HIRESb spectra from three epochs as well as with the HIRESr spectrum in the wavelength region where the b and r spectra overlap (see Table~\ref{tab1:obs}). No dramatic changes in line shape or radial velocity of the sort reported in Figure~5 of \citet{xu_shallow_2019} in WD\,1145+017 are apparent in WD\,J0234-0406. With the caveat that the circumstellar lines are typically broad and weak and often blended with other circumstellar lines or photospheric lines, no variations in radial velocity or EW are noticeable.

That said, secular variability of the sort apparent in WD\,1145+017 can come and go \citep{aungwerojwit_long-term_2024}. Thus, it is possible, albeit unlikely, that short term variations may someday be seen in WD\,J0234-0406.

\subsection{Gaseous Circumstellar Features}\label{sec4p3}

\begin{figure}
\centering
\includegraphics[width=\linewidth,trim={0.8cm 6.8cm 1.5cm 8.0cm},clip]{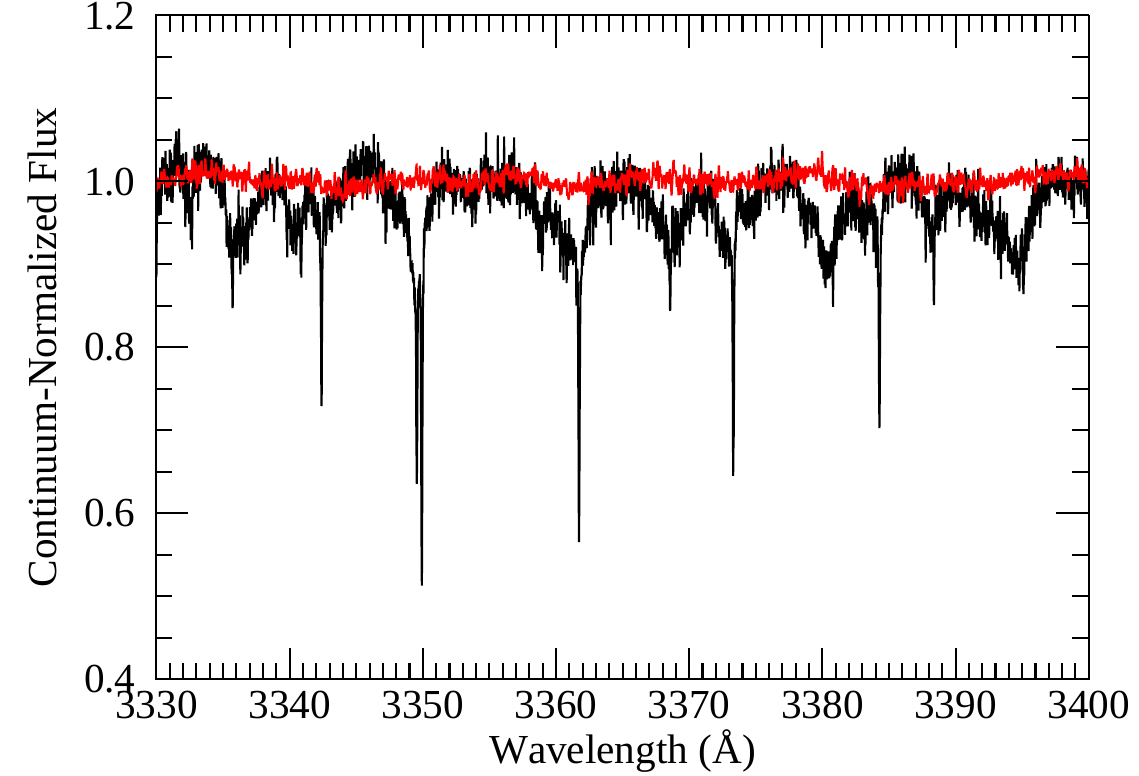}
\caption{WD\,J0234-0406 (black) vs the featureless white dwarf calibration star EGGR 180 (red). The deep narrow features are photospheric. The shallower, broader, features are circumstellar. Identification of a few prominent lines: 3349.03 and 3349.40 doublet, 3361.21, and 3272.79 \ion{Ti}{2}. 3368.04 \ion{Cr}{2}. 3380.11 \ion{Fe}{1}. 3390 - 3396 is a blend of lines (see Table~\ref{tab4:simple}). \label{fig4:wd-vs-calib}}
\end{figure}

\begin{figure}
\centering
\includegraphics[width=\linewidth,trim={0.9cm 6.8cm 1.5cm 8.0cm},clip]{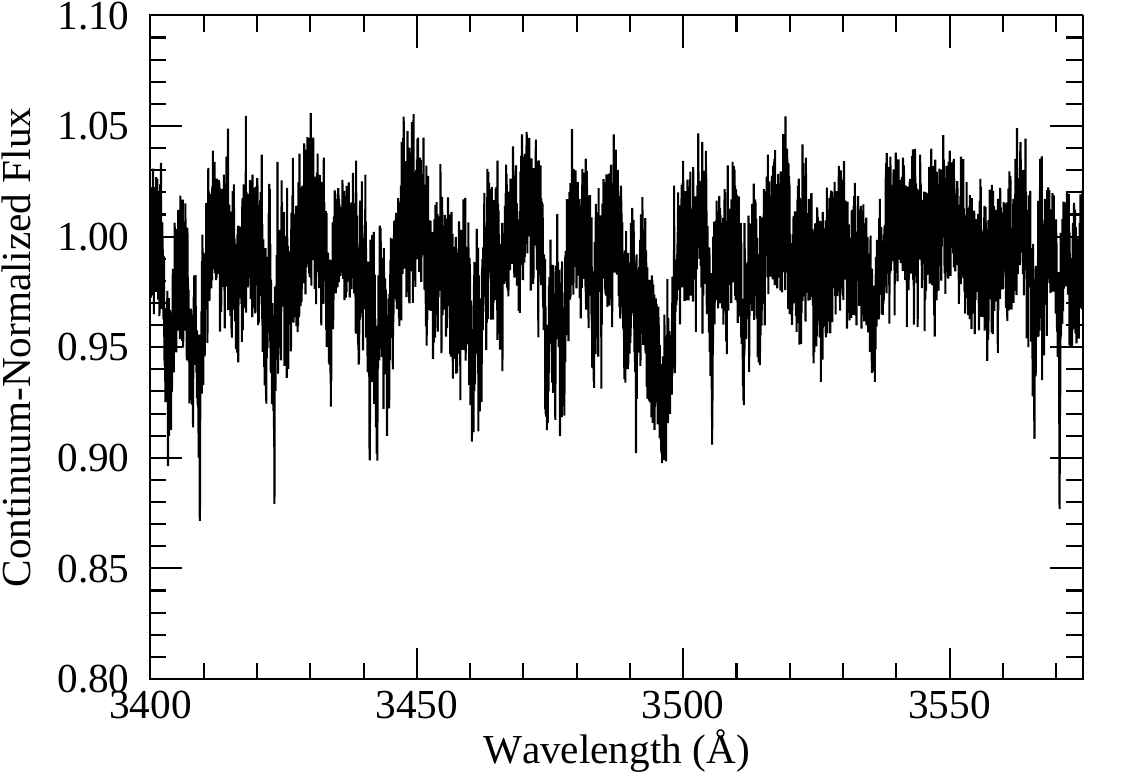}
\caption{Spectral region dominated by circumstellar absorption features.\label{fig5:csabs}}
\end{figure}

\begin{figure}
\centering
\includegraphics[width=\linewidth]{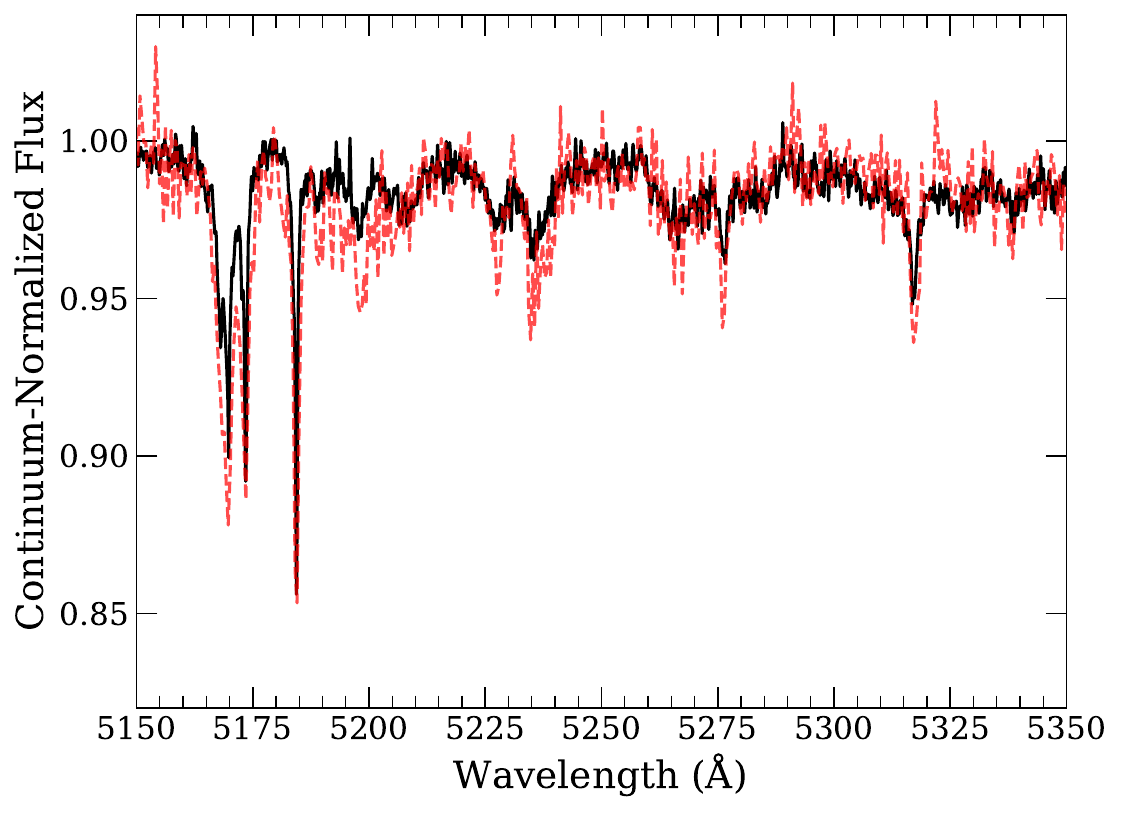}
\caption{Representative HIRES (black), smoothed to match MagE spectral resolution, and MagE (red) spectra showing generally good agreement. \label{fig6:comp_hires-mage}}
\end{figure}

Absorption features from the circumstellar gas are distinguished from photospheric lines by the much narrower line widths and somewhat more positive radial velocities of the latter. Some optical transitions display broad/shallow gaseous plus deep/narrow photospheric components, others show only one component (e.g., Figure~\ref{fig4:wd-vs-calib}). Figure~\ref{fig5:csabs} displays a spectral region dominated by circumstellar gas absorption features. Figure~\ref{fig6:comp_hires-mage} is a comparison between HIRES and MagE spectra. Figures~\ref{fig7:fits01} to \ref{fig10:disk-vs-no-disk} illustrate our circumstellar model fits to HIRES and COS spectra of WD\,J0234-0406. The COS resolution is insufficient to resolve the circumstellar and photospheric components. 

\begin{figure}
\centering
\includegraphics[width=\linewidth,trim={1.0cm 6.8cm 1.5cm 8.0cm},clip]{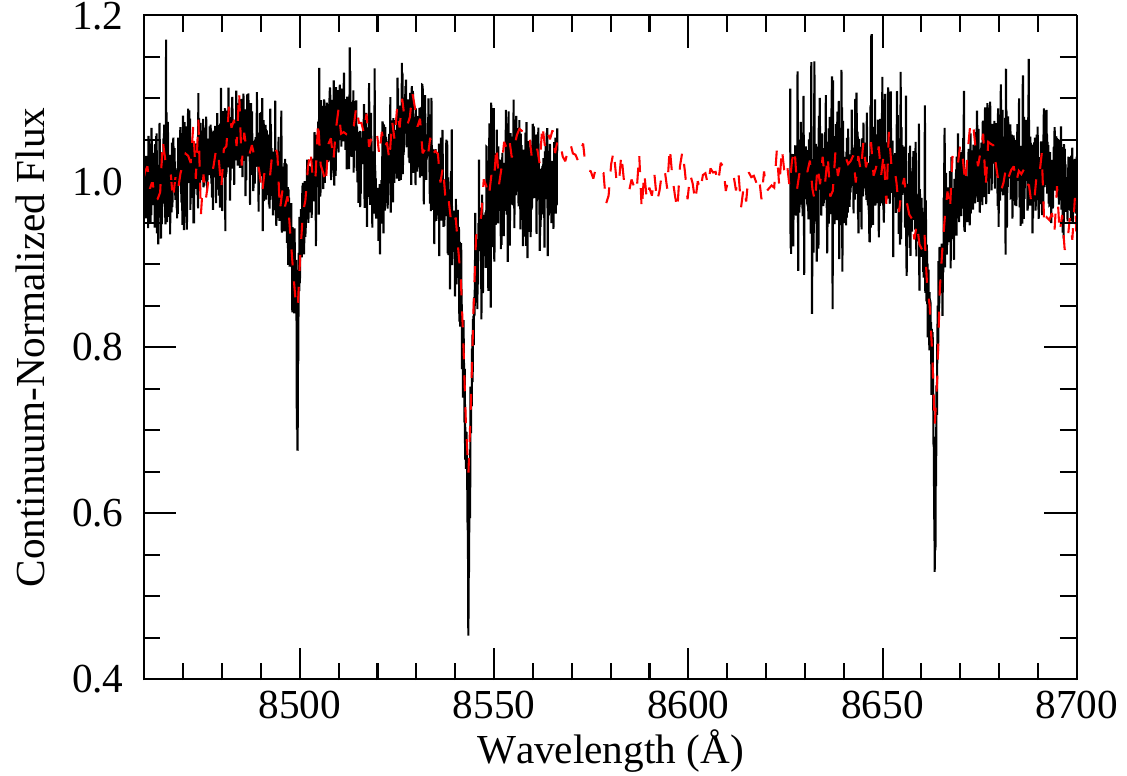}
\caption{Weak Ca IR triplet emission (mentioned by GF21) as measured with HIRES (black) and MagE (red). The dip in the HIRES spectrum at 8520 {\AA} is due to a bad column in the red HIRES detector and not an absorption feature in the white dwarf spectrum. \label{fig11:ca-triplet}}
\end{figure}

GF21 found a few weak emission lines. Specifically, they state: ``The only emission features from the gas disc around WD\,J0234-0406 are the \ion{Ca}{2} infrared triplet and the indistinguishable blend of \ion{Fe}{2}/\ion{Mg}{1} lines near 5175 \AA''. There is no mention in GF21 of the broad absorption features that are the focus of this paper. We suspect that this might have been due, at least in part, to an inadequate spectral resolution that ranged between 4000 and 8000. Figure~\ref{fig11:ca-triplet} displays the weak Ca IR triplet emission (mentioned by GF21) as measured with HIRES and MagE.

GF21 note that WD\,J0234-0406 is the ``first white dwarf of spectral type DABZ found to host a gaseous debris disc." See Section~\ref{sec4p4} for consideration of the spectral classification of WD\,J0234-0406.

Table~\ref{tab3:lines} presents a list of circumstellar lines present and perhaps present in the spectrum of WD\,J0234-0406. Because of the broad line widths it is sometimes not possible to decide which of two elements, or perhaps both, is responsible for a given spectral feature. In such cases, transitions of both elements are listed and marked with an asterisk.

In regard to \ion{Mg}{1}, photospheric fits to transitions at 3832.30 and 3838.292 {\AA} indicate excess absorption in the blue wing of the lines, likely consistent with the longer wavelength circumstellar \ion{Mg}{1} lines listed in Table~\ref{tab3:lines}.

In some portions of the spectra so many gaseous lines are present that they blend together and depress the continuum over a broad range of wavelengths. The most obvious of these wavelength ranges are listed in Table~\ref{tab4:simple}. As noted in the table footnote, because laboratory conditions do not match those in the circumstellar gas of WD\,J0234-0406, the dominant ion identifications given in the table should be regarded as suggestive, rather than definitive. Examples of regions of depressed continuum can be seen in Figures~\ref{fig4:wd-vs-calib} to \ref{fig6:comp_hires-mage}.

\begin{figure}
\centering
\includegraphics[width=\linewidth,trim={1.0cm 6.8cm 1.5cm 8.0cm},clip]{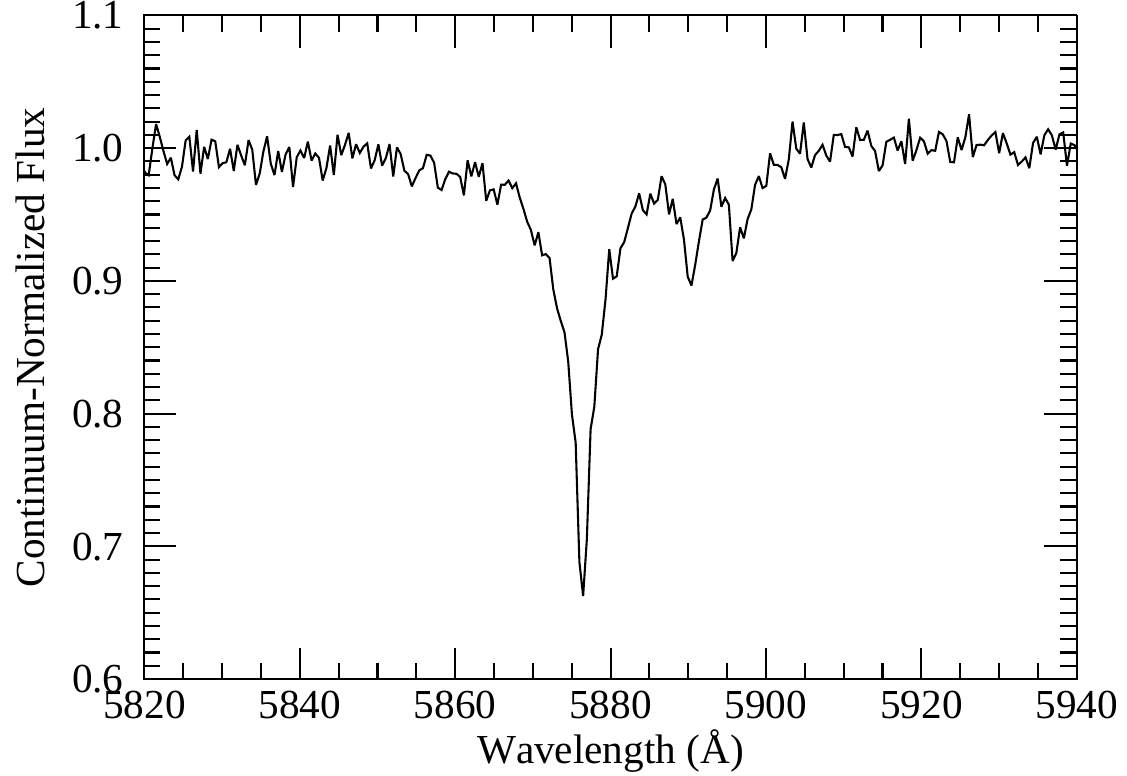}
\caption{Sodium D-lines (5890 and 5896 \AA) measured with MagE.\label{fig12:sodium}}
\end{figure}

A striking aspect of the spectrum of WD\,J0234-0406 are the Na D1 and D2 lines near 5900 \AA, with equivalent width (EW) of 210 and 130 m\AA, respectively (Table~\ref{tab3:lines} and Figure~\ref{fig12:sodium}). \citet{vanderbosch_recurring_2021} detect these Na lines with comparable EW, but an order of magnitude smaller linewidths, at white dwarf ZTF J0328-1219. They argue that the Na lines likely originate in circumstellar gas.

\clearpage
\startlongtable
\begin{deluxetable*}{cccccccc}

\tablecaption{Circumstellar Lines \label{tab3:lines}}
\tablehead{
\colhead{Ion} & \colhead{$\lambda_{o}$} & \colhead{$\lambda$} & \colhead{RV} & \colhead{E} & \colhead{$\log gf$} & \colhead{EW} & \colhead{FWHM} \\[-4pt]
\colhead{} & \colhead{(\AA)} & \colhead{(\AA)} & \colhead{(km s$^{-1}$)} & \colhead{(eV)} & \colhead{} & \colhead{(m\AA)} & \colhead{(\AA)} 
}

\startdata
\ion{Ca}{1}&4226.728&4226.83&7.2&0&0.24&220&2.9\\
\ion{Ca}{1} & 4454.779 & 4455.31 & 35.8 & 1.9 & 0.24 & 30 & 0.8 \\
\ion{Ca}{2} & 8498.02 & 8499.09 & 37.8 & 1.7 & -1.26 & 1050 & 6.1 \\
\ion{Ca}{2} & 8542.09 & 8542.90 & 28.4 & 1.7 & -0.31 & 1650 & 6.3 \\
\ion{Ca}{2} & 8662.14 & 8662.75 & 21.1 & 1.7 & -0.57 & 1100 & 5.4 \\
\ion{Cr}{2} & 3118.646 & 3118.90 & 24.4 & 2.4 & 0.00 & 130 & 0.9 \\
\ion{Cr}{2} & 3124.973* & 3125.43 & 43.9 & 2.4 & \nodata &  140 & 0.5 \\
\ion{Cr}{2} & 3368.05 & 3368.61 & 49.9 & 2.5 & -0.09 & 250 & 2.0 \\
\ion{Cr}{2} & 3403.307* & 3403.60 & 25.8 & 2.4 & -0.67 & 180 & 1.8 \\
\ion{Cr}{2} & 3422.732 & 3423.20 & 48.0 & 2.5 & -0.41 & 200 & 1.7 \\
\ion{Cr}{2} & 3433.295 & 3433.66 & 31.9 & 2.4 & -0.73 & 120 & 1.4 \\
\ion{Fe}{1}? & 3124.885* & 3125.43 & 52.3 & 3.2 & -0.97 & 140 & 0.5 \\
\ion{Fe}{1} & 3136.50 & 3136.99 & 46.9 & 3.3 & -0.60 & 50 & 0.7 \\
\ion{Fe}{1} & 3196.927 & 3197.34 & 38.8 & 2.4 & 0.14 & 140 & 1.0 \\
\ion{Fe}{1} & 3254.362* & 3254.68 & 29.3 & 3.3 & -0.06 & 50 & 0.35 \\ 
\ion{Fe}{1} & 3380.110$\ast$ & 3380.41 & 26.6 & 2.8 & -0.70 & 260 & 1.6 \\
\ion{Fe}{1} & 3403.290* & 3403.60 & 27.3 & 2.8 & -1.51 & 180 & 1.8 \\
\ion{Fe}{1} & 3443.876 & 3444.34 & 40.4 & 0.1 & -1.37 & 60 & 1.6 \\
\ion{Fe}{1} & 3459.914 & 3460.45 & 46.5 & 3.0 & -0.88 & 150 & 1.7 \\
\ion{Fe}{1} & 3513.818 & 3514.22 & 34.3 & 0.9 & -1.16 & 100 & 1.7 \\
\ion{Fe}{1} & 3565.379 & 3566.00 & 52.2 & 1.0 & -0.13 & 170 & 2.0 \\
\ion{Fe}{1} & 3570.01 & 3570.46 & 37.8 & 2.4 & -1.19 & 40 & 0.8 \\
\ion{Fe}{1} & 3581.193* & 3581.31 & 10.0 & 0.9 & 0.41 & 35 & 0.6 \\
\ion{Fe}{1} & 3585.319 & 3585.86 & 45.3: & 0.95 & -0.80 & 140 & 1.6 \\
\ion{Fe}{1}? & 3593.322* & 3593.78 & 38.2 & 3.3 & \nodata &  80 & 1.1 \\ 
\ion{Fe}{1} & 3613.44 & 3614.03 & 49.0 & 3.3 & -0.76 & 90 & 1.4 \\
\ion{Fe}{1} & 3618.768 & 3619.29 & 43.3 & 1.0 & 0.0 & 80 & 1.8 \\
\ion{Fe}{1} & 3631.097 & 3631.43 & 27.5 & 2.8 & -0.18 & 150 & 1.9 \\ 
\ion{Fe}{1} & 3651.467 & 3651.92 & 37.2 & 2.8 & 0.02 & 20 & 0.55 \\
\ion{Fe}{1} & 3677.628 & 3678.29 & 54 & 2.8 & -0.21 & 130 & 1.6 \\ 
\ion{Fe}{1} & 3694.006 & 3694.42 & 33.6 & 3.0 & 0.25 & 30 & 0.6 \\ 
\ion{Fe}{1}? & 3814.523* & 3815.18 & 51.7 & 1.0 & -2.39 & 180 & 3.8 \\ 
\ion{Fe}{1} & 3820.425 & 3820.84 & 32.6 & 0.9 & 0.12 & 120 & 2.8 \\
\ion{Fe}{1} & 3859.911 & 3860.13 & 17 & 0 & -0.71 & 40 & 0.9 \\
\ion{Fe}{1} & 4132.533* & 4133.13 & 43.3 & 4.3 & -0.86 & 80 & 2.0 \\
\ion{Fe}{1} & 4132.899* & 4133.13 & 16.8 & 2.8 & -1.0 & 80 & 2.0 \\
\ion{Fe}{1} & 4173.315* & 4173.76 & 32.0 & 2.8 & -1.76 & 150 & 3.1 \\
\ion{Fe}{1}? & 4215.423* & 4215.94 & 36.8 & 3.0 & -1.76 & 120 & 2.0 \\
\ion{Fe}{1} & 4289.915* & 4290.33 & 29.0 & 3.4 & -1.54 & 180 & 3.0 \\ 
\ion{Fe}{1} & 4443.194* & 4443.97 & 52.2 & 2.9 & -0.97 & 160 & 3.1 \\
\ion{Fe}{1} & 5167.488* & 5168.05 & 32.6 & 1.5 & -1.12 & 20 & 0.4 \\
\ion{Fe}{1} & 5586.756 & 5587.53 & 41.6 & 3.4 & 0.10 & 250 & 7: \\
\ion{Fe}{2} & 3135.362 & 3135.75 & 37.1 & 3.9 & -1.1 & 120 & 0.9 \\
\ion{Fe}{2} & 3183.11 & 3183.56 & 42.4 & 1.7 & -2.05 & 90 & 0.8 \\
\ion{Fe}{2} & 3255.886 & 3256.29 & 37.2 & 1.0 & -2.45 & 50 & 0.5 \\
\ion{Fe}{2} & 3258.773 & 3259.26 & 44.8 & 3.9 & -0.92 & 100 & 0.6 \\
\ion{Fe}{2}? & 3295.819* & 3296.1 & 25.6 & 1.1 & -2.9 & 180 & 1.8 \\
\ion{Fe}{2} & 4173.451* & 4173.76 & 22.2 & 2.6 & -2.16 & 150 & 3.1 \\ 
\ion{Fe}{2} & 4233.162 & 4233.70 & 38.1 & 2.6 & -1.97 & 70 & 0.9 \\
\ion{Fe}{2} & 4385.377* & 4385.86 & 33.0 & 2.8 & -2.59 & 340 & 3.3 \\
\ion{Fe}{2} & 4416.819* & 4417.20 & 25.9 & 2.8 & -2.61 & 240 & 3.8 \\
\ion{Fe}{2} & 4508.28 & 4508.77 & 32.6 & 2.8 & -2.37 & 70 & 1.4 \\
\ion{Fe}{2} & 4522.628 & 4523.03 & 26.7 & 2.8 & -2.25 & 50 & 1.0 \\ 
\ion{Fe}{2} & 4549.466$\ast$ & 4550.04 & 37.9 & 2.9 & -1.73 & 180 & 1.4 \\ 
\ion{Fe}{2} & 4583.829* & 4584.18 & 23.0 & 2.8 & -1.74 & 180 & 2.0 \\
\ion{Fe}{2} & 4923.921* & 4924.69 & 46.8 & 2.9 & -1.2 & 340 & 3.9 \\
\ion{Fe}{2} & 5018.436 & 5018.86 & 25.3 & 2.9 & -1.1 & 50: & 0.5: \\
\ion{Fe}{2} & 5169.028 & 5169.4 & 21.6 & 2.9 & -1.0 & 570 & 5.3 \\
\ion{Fe}{2} & 5316.609 & 5317.14 & 30.5 & 3.1 & -1.87 & 140 & 2.0 \\
\ion{Mg}{1} & 5167.321* & 5168.05 & 42.3 & 2.7 & -0.87 & 20 & 0.4 \\
\ion{Mg}{1} & 5172.68 & 5173.25 & 33.0 & 2.7 & -0.45 & 60 & 1.7 \\
\ion{Mg}{1} & 5183.604 & 5183.94 & 19.4 & 2.7 & -0.18 & 60 & 1.6 \\
\ion{Mn}{2} & 3441.988 & 3442.45 & 40.2 & 1.8 & -0.35 & 80 & 0.8 \\
\ion{Mn}{2} & 3474.1* & 3474.52 & 36.3 & 1.8 & -0.93 & 100 & 1.1 \\
\ion{Mn}{2} & 3482.905 & 3483.19 & 24.5 & 1.8 & -0.84 & 70 & 1.0 \\
\ion{Na}{1} & 5889.950 & 5890.46 & 26.0 & 0 & 0.11 & 400** & 3.0** \\
\ion{Na}{1} & 5895.924 & 5896.32 & 20.2 & 0 & -0.19 & 320** & 3.0** \\
\ion{O}{1} & $\sim$7774 & 7774.66 & Blend & 9.1 & $\sim$0.2 & $\sim$700 & $\sim$6 \\
\ion{Ti}{2}& 3162 & Blend of 3 & \nodata & 0.12 & \nodata & 300 & 1.9 \\
\ion{Ti}{2} & 3190.879 & 3191.09 & 19.8 & 1.1 & 0.23 & 20 & 0.25 \\
\ion{Ti}{2} & 3217.063 & 3217.43 & 34.2 & 0.027 & -0.49 & 260 & 1.7 \\
\ion{Ti}{2} & 3222.843 & 3223.00 & 14.6 & 0.01 & -0.42 & 70 & 1.0 \\
\ion{Ti}{2} & 3229.198 & 3229.53 & 30.8 & 0 & -0.56 & 190 & 1.1 \\
\ion{Ti}{2} & 3254.247* & 3254.68 & 39.9 & 0.05 & -0.56 & 50 & 0.35 \\
\ion{Ti}{2} & 3261.585 & 3261.80 & 19.8 & 1.89 & 0.53 & 30 & 0.4 \\
\ion{Ti}{2} & 3271.655 & 3271.98 & 29.8 & 1.24 & -0.28 & 160 & 1.8 \\
\ion{Ti}{2} & 3276.772 & 3277.41 & 58.4 & 1.18 & -0.89 & 210 & 2.3 \\
\ion{Ti}{2} & 3287.661 & 3288.25 & 53.7 & 1.9 & 0.46 & 220 & 1.5 \\
\ion{Ti}{2} & 3332.111 & 3332.55 & 39.5 & 1.2 & -0.11 & 60 & 0.6 \\
\ion{Ti}{2} & 3349.2 & Blend of 2 & \nodata & \nodata & \nodata & 200 & \nodata\\
\ion{Ti}{2} & 3372.80 & 3373.0 & 18.0  & 0.01  & 0.28  & 110 & 1.3  \\
\ion{Ti}{2} & 3380.279$\ast$ & 3380.41 & 11.6 & 0.05 & -0.63 & 260 & 1.6 \\
\ion{Ti}{2} & 3387.846 & 3388.29 & 39.3 & 0.03 & -0.41 & 120 & 1.1 \\
\ion{Ti}{2} & 3407.20 & 3407.7 & 43.1 & 0.05 & -2.0 & 120 & 1.3 \\
\ion{Ti}{2} & 3510.845 & 3511.33 & 41.4 & 1.9 & 0.29 & 70 & 0.9 \\
\ion{Ti}{2}? & 3814.58* & 3815.18 & 47.2 & 0.6 & -1.61 & 180 & 3.8 \\ 
\ion{Ti}{2} & 3900.551 & 3901.09 & 41.4 & 1.1 & -0.20 & 25 & 0.5 \\
\ion{Ti}{2} & 3913.465 & 3913.95 & 37.2 & 1.1 & -0.42 & 60 & 1.2 \\
\ion{Ti}{2} & 4395.033 & 4395.57 & 36.6 & 1.1 & -0.66 & 60 & 1.0 \\
\ion{Ti}{2} & 4468.50 & 4469.08 & 38.9 & 1.1 & -0.60 & 40 & 0.7 \\
\ion{Ti}{2} & 4501.270 & 4501.69 & 28.0 & 1.1 & -0.77 & 60 & 1.1 \\ 
\ion{Ti}{2} & 4533.969 & 4534.60 & 41.8 & 1.2 & -0.77 & 80 & 1.3 \\
\ion{Ti}{2} & 4549.617$\ast$ & 4550.04 & 27.9 & 1.6 & -0.10 & 180 & 1.4 \\
\ion{Ti}{2} & 4571.968 & 4572.62 & 42.8 & 1.6 & -0.32 & 100 & 1.8 \\
\ion{V}{1}? & 3295.462* & 3296.1 & 58.1 & 1.4 & -1.92 & 180 & 1.8 \\
\ion{V}{1}? & 3295.78* & 3296.1 & 29.1 & 1.2 & \nodata & 180 & 1.8 \\
\ion{V}{1} & 4135.296 & 4135.73 & 31.5 & 1.1 & -2.0 & 50 & 1.2 \\
\ion{V}{1} & 4385.327* & 4385.86 & 36.5 & 2.1 & -1.0 & 340 & 3.3 \\
\ion{V}{1} & 4416.468* & 4417.20 & 49.7 & 0.27 & -0.83 & 240 & 3.8 \\
\ion{V}{1} & 4434.594 & 4435.05 & 30.8 & 1.9 & -0.34 & 100: & 3.0: \\
\ion{V}{1} & 4443.336* & 4443.97 & 42.8 & 2.7 & 0.36 & 160 & 3.1 \\
\ion{V}{1}? & 4583.78* & 4584.18 & 26.2 & 1.9 & \nodata & 180 & 2.0 \\
\ion{V}{1}? & 5275.6 & 5276.08 & 27.3 & 2.3 & \nodata & 140 & 2.9 \\ 
\ion{Sc}{2} & 3572.526 & 3572.80 & 23.0 & 0.02 & 0.27 & 60 & 0.8 \\
\ion{Sc}{2} & 3576.34 & 3576.83 & 41.1 & 0 & 0.01 & 50 & 1.2 \\
\ion{Sc}{2} & 3580.925* & 3581.31 & 32.3 & 0 & -0.15 & 35 & 0.6 \\
\ion{Sc}{2} & 4246.822 & 4247.16 & 23.9 & 0.3 & 0.24 & 20 & 0.4 \\
\ion{Sr}{2} & 4077.71 & 4078.1 & 28.7 & 0 & 0.15 & 80 & 1.1 \\
\ion{Sr}{2} & 4215.524* & 4215.95 & 30.3 & 0 & -0.17 & 120 & 2.0\\
\enddata
\tablecomments{Air rest wavelengths $\lambda_o$ \citep{NIST_ASD,van_hoof_recent_2018}; $\lambda$ measured wavelength; RV radial velocity with respect to the Sun; E lower state energy; EW equivalent width; FWHM full width at half maximum depth. $\ast$ indicates line blended with another listed element. \ion{Fe}{1} at 3378.679 and 3402.256 appear to be present as does some blend of \ion{Fe}{1} 3475.450, 3475.651 $\&$ 3476.344. The \ion{Fe}{1} line at 3820.425 \AA\ might be blended with photospheric \ion{He}{1} at 3819.61 \AA. The \ion{Fe}{2} line at 4923.921 might be blended with photospheric \ion{He}{1} at 4921.93. For \ion{Na}{1}, double $\ast$ indicates that the EW and FWHM entries are obtained from the MagE spectrum and not from a HIRES spectrum.}
\end{deluxetable*}

\begin{deluxetable}{cll}[h]
\tablecaption{Regions of Broad Circumstellar Absorption in WDJ0234 \label{tab4:simple}}
\tablehead{
\colhead{Wavelength range (\AA)} & \colhead{Dominant Ions} & \colhead{Other Ions}
}
\startdata
3145--3150 & \ion{Fe}{1} & \ion{V}{1} \\
3175--3200 & \ion{Fe}{1}, \ion{Fe}{2} & \ion{Ti}{2}, \ion{V}{1} \\
3208--3220 & \ion{Fe}{1} & \ion{Fe}{2}, \ion{Ti}{2}, \ion{V}{1} \\
3220--3246 & \ion{Fe}{1}, \ion{Ti}{2} & \ion{Fe}{2}, \ion{V}{1} \\
3300--3320 & \ion{Fe}{1} & \ion{V}{1} \\
3320--3330 & \ion{Fe}{1} & \ion{Ti}{2}, \ion{V}{1} \\
3335--3345 & \ion{Fe}{1}, \ion{Ti}{2}, \ion{V}{1} & \nodata\\
3358--3362 & \ion{Ti}{2} & \ion{Fe}{1}, \ion{Sc}{2}? \\ 
3378--3385 & \ion{Fe}{1}, \ion{Ti}{2} & \nodata\\
3390--3396 & \ion{Fe}{1}, \ion{Ti}{2}, \ion{V}{1} & \nodata\\
3450--3470 & \ion{Fe}{1}, \ion{V}{1} & \ion{Ti}{2} \\
3641--3652 & \ion{Fe}{1}, \ion{V}{1} & \nodata\\
3745--3751 & \ion{Fe}{1}, \ion{V}{1} & \nodata\\
3756--3763 & \ion{Fe}{1} & \ion{Ti}{2}, \ion{V}{1} \\
3784--3802 & \ion{Fe}{1}, \ion{V}{1} & \nodata\\
3823--3847 & \ion{Mg}{1} & \ion{Fe}{1}, \ion{V}{1} \\
5203--5211 & \ion{Fe}{1}, \ion{V}{1} & \nodata\\
\hline
\enddata
\tablecomments{The contributions of the listed ions to the regions where broad absorption of the continuum is noted is based on laboratory data given in \citet{NIST_ASD} and \citet{van_hoof_recent_2018}. Because the laboratory conditions do not match those in the circumstellar gas of WD\,J0234-0406, the dominant ion identifications given here should be regarded as suggestive, rather than definitive. \ion{Ca}{1} (circumstellar) might contribute in some wavelength ranges.}
\end{deluxetable}

The Na lines at WD\,J0234-0406 are unlikely to be photospheric due to the high temperature of the star. In addition, the radial velocities of the two lines differ substantially from the velocity of the photosphere (see Section~\ref{sec5}). Thus, the Na lines are either circumstellar or interstellar, but very probably the former for the following reasons. Table 1 in \citet{sfeir_mapping_1999} lists 144 stars between 50 and 200+ pc from Earth. WD\,J0234-0406 is 92 pc from Earth. Of the 144 stars, 62 are within 100 pc -- none have D2 EW as large as 120 m\AA. Of the stars beyond 100 pc a few have EW $>$120 m\AA; the most extreme example by far is HD\,188350 at 106 pc and with D2 EW = 218 m\AA. In most directions, the local low-density interstellar solar bubble extends out to $<$100 pc. The stars closest to WD\,J0234-0406 in the D. M. Sfier et al. table are HD 18543 and HD 18633. These are $>$100 pc from Earth ($pi$ = 8.94 and 8.61 mas) and about 8 deg away from WD\,J0234-0406 in the
plane of the sky. Both have D2 EW $<$3.5~m\AA.

Another, even stronger, reason why the Na lines are circumstellar is their line widths (FWHM in Table~\ref{tab3:lines}) which are far larger than widths of interstellar Na lines. \citet{hobbs_observations_1978} presents data for interstellar Na absorption lines toward 44 stars; the line widths are $\leq$20~km~s$^{-1}$.

\subsection{Photospheric Spectrum and Stellar Classification}\label{sec4p4}

The optical spectrum of WD\,J0234-0406 contains strong photospheric Balmer lines of hydrogen -- see Figure~\ref{fig1:fit_photo} above and Figure~2 in GF21. WD\,J0234-0406 stands out as extremely H-rich among He-dominated white dwarfs with similar effective temperatures \citep[see fig. 5 of][]{rolland_spectral_2018}. WD\,J0234-0406 also displays photospheric neutral helium lines at 4471.50, 5015.68, 5875.70, and 6678.15 \AA. \ion{He}{1} 3888.65\,{\AA} may be present but it is blended with H $\zeta$ at 3889.05\,{\AA}. \ion{He}{1} 3819.61\,{\AA} may be present but blended with a circumstellar line of \ion{Fe}{1}, and \ion{He}{1} 4921.93\,{\AA} may be present but blended with a circumstellar \ion{Fe}{2} line (see note to Table~\ref{tab3:lines}). A strong (EW $\sim$ 1 \AA), broad, line at about 4715 $\AA$ may be \ion{He}{1} 4713.2\,{\AA}. Even though the He lines are weaker than the H lines, the white dwarf is actually helium-rich, as first noted by GF21. 

\begin{deluxetable*}{cl}[ht]
    \tablecaption{Optical Photospheric Lines \label{tab5:optical}}
    \tablehead{\colhead{Ion} & \colhead{Air Wavelength (\AA)}}
    \startdata
    \ion{Al}{1} & 3944.00, 3961.52 \\
\ion{Ca}{1} & 4226.73 \\
\ion{Ca}{2} & 3158.87, 3179.33, 3181.78, 3706.02, 3736.90, 3933.66, 3968.47, 8498.02, 8542.09, 8662.14 \\
\ion{Cr}{2} & 3124.97b, 3132.05, 3368.04, 3408.76, 3422.73, 3433.30 \\
\ion{Fe}{1} & 3124.89b, 3392.62, 3394.58b, 3570.10, 3581.19, 3608.86, 3618.77, 3677.63, 3719.93, 3734.86, 3749.49, 3758.23, 3759.29,\\
& 3761.32, 3763.79, 3767.19, 3815.84, 3820.43, 3825.88, 3827.82, 3834.22, 3855.85b, 3859.91, 4063.59, 4071.74, 4471.70b,\\
& 5167.49b, 6820.37? \\
\ion{Fe}{2} & 3154.20b, 3167.86, 3193.80, 3196.07, 3210.44, 3213.31, 3227.74, 3258.77, 5018.436, 5169.028 \\
\ion{Mg}{1} & 3829.36, 3832.30, 3838.29, 5167.32b, 5172.68, 5183.60 \\
\ion{Mg}{2} & 4481.13, 4481.33 \\
\ion{O}{1} & 7771.94, 7774.17, 7775.39 \\ 
\ion{Ti}{2} & 3154.19b, 3161.77, 3162.57, 3168.52, 3190.88, 3202.53, 3217.05, 3222.84, 3224.24, 3229.19, 3232.28, 3234.51, 3236.57,\\
& 3239.04, 3241.98, 3248.60, 3251.91, 3252.91, 3254.25, 3261.58, 3271.66, 3287.66, 3322.93, 3329.45, 3335.19, 3340.34,\\
& 3341.87, 3346.75, 3349.03, 3349.40, 3361.21, 3372.79, 3380.28, 3383.76, 3387.83, 3394.58b, 3685.20 \\
\ion{V}{1}?? & 3855.84b \\
\enddata
\tablecomments{Rest Wavelengths: if followed by a "b" indicate a potential blend of two listed elements. See Section~\ref{sec4p4} for a discussion of photospheric lines of \ion{He}{1}.}
\end{deluxetable*}

\begin{deluxetable*}{cl}[ht]
    \tablecaption{Ultraviolet Photospheric Lines \label{tab6:uv}}
    \tablehead{\colhead{Ion} & \colhead{Vaccuum Wavelength (\AA)}}
    \startdata
    \ion{Al}{2} & 1189.19, 1190.05, 1191.81, 1211.90 \\
\ion{C}{2} & 1334.53, 1335.71 \\
\ion{Fe}{2} & 1112.05, 1121.98, 1122.53, 1128.72, 1138.63, 1142.31, 1143.23, 1144.94, 1147.41, 1148.28, 1151.15, 1169.19, 1175.68, \\
& 1183.83, 1198.93, 1224.13, 1230.93, 1260.53, 1266.68, 1267.42,  1271.98, 1272.61, 1275.78, 1277.64, 1296.70, 1358.94,\\
& 1361.38, 1368.09, 1371.02, 1375.17, 1379.47, 1392.82, 1408.48, 1412.84, 1424.72 \\
\ion{Mg}{2} & 1239.93, 1240.40 \\
\ion{Ni}{2} & 1133.73, 1134.53, 1137.09, 1154.42, 1164.58, 1168.04, 1171.29, 1317.22, 1335.20, 1370.13, 1381.29, 1411.07 \\
\ion{O}{1} & 1152.15, 1217.65, 1302.17, 1304.86, 1306.03 \\
\ion{Si}{1} & 1258.80 \\
\ion{Si}{2} & 1190.42, 1193.29, 1194.50, 1197.39, 1223.90, 1224.25, 1226.80, 1226.98, 1227.60, 1228.61, 1228.74, 1229.34, 1246.74,\\
& 1248.43, 1250.09, 1240.44, 1251.16, 1260.42, 1264.74, 1265.00, 1304.37, 1309.28, 1346.88, 1348.54, 1350.07, 1352.64,\\
& 1353.72\\
\ion{Si}{3} & 1113.23, 1206.5, 1294.55, 1296.73, 1298.95 \\
\enddata
\end{deluxetable*}

\begin{figure*}[ht]
\includegraphics[trim={1cm 9.5cm 1.8cm 5.5cm},clip,width=\linewidth]{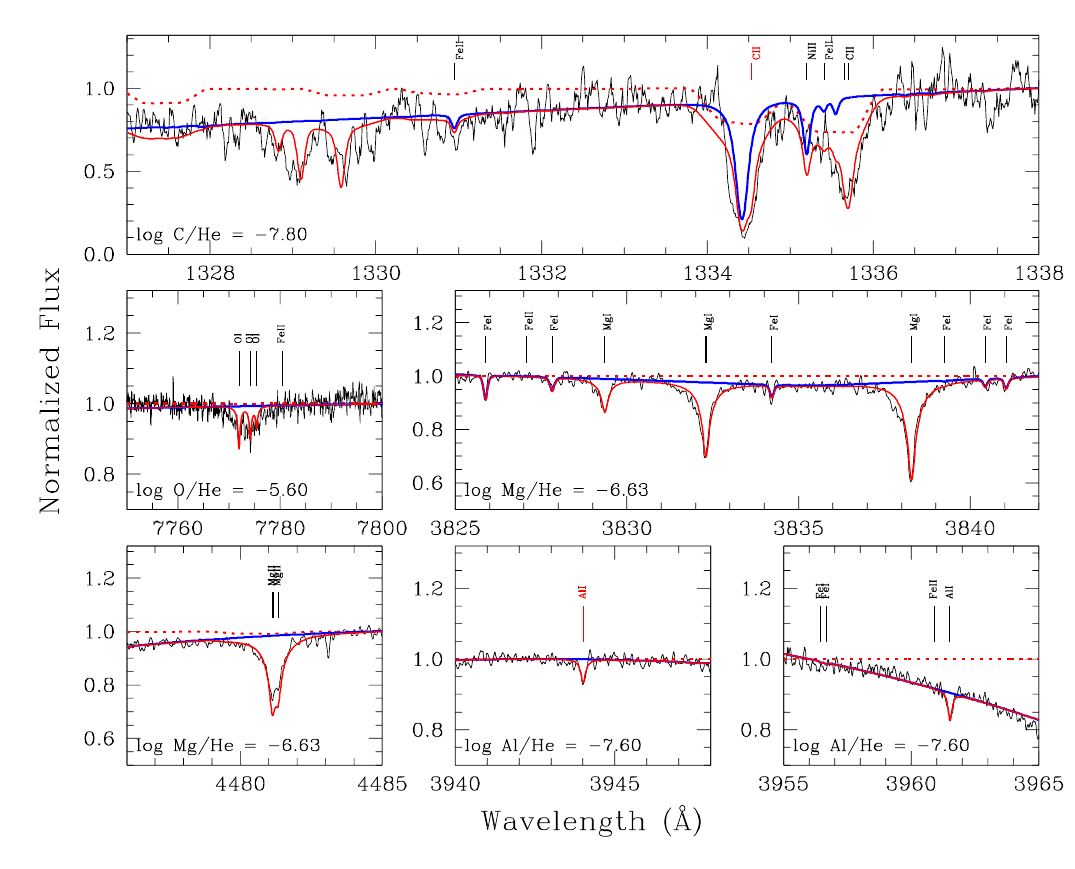}
\caption{Selected regions of the photospheric fit (red) to WD\,J0234-0406 and its circumstellar disk (dotted red). The model photospheric + circumstellar gas spectrum without the fitted photospheric element is shown in blue. A red label is used for lines coming from the ground state.\label{fig7:fits01}}
\end{figure*}

\begin{figure*}[ht]
\includegraphics[trim={1cm 9.5cm 1.8cm 5.5cm},clip,width=\linewidth]{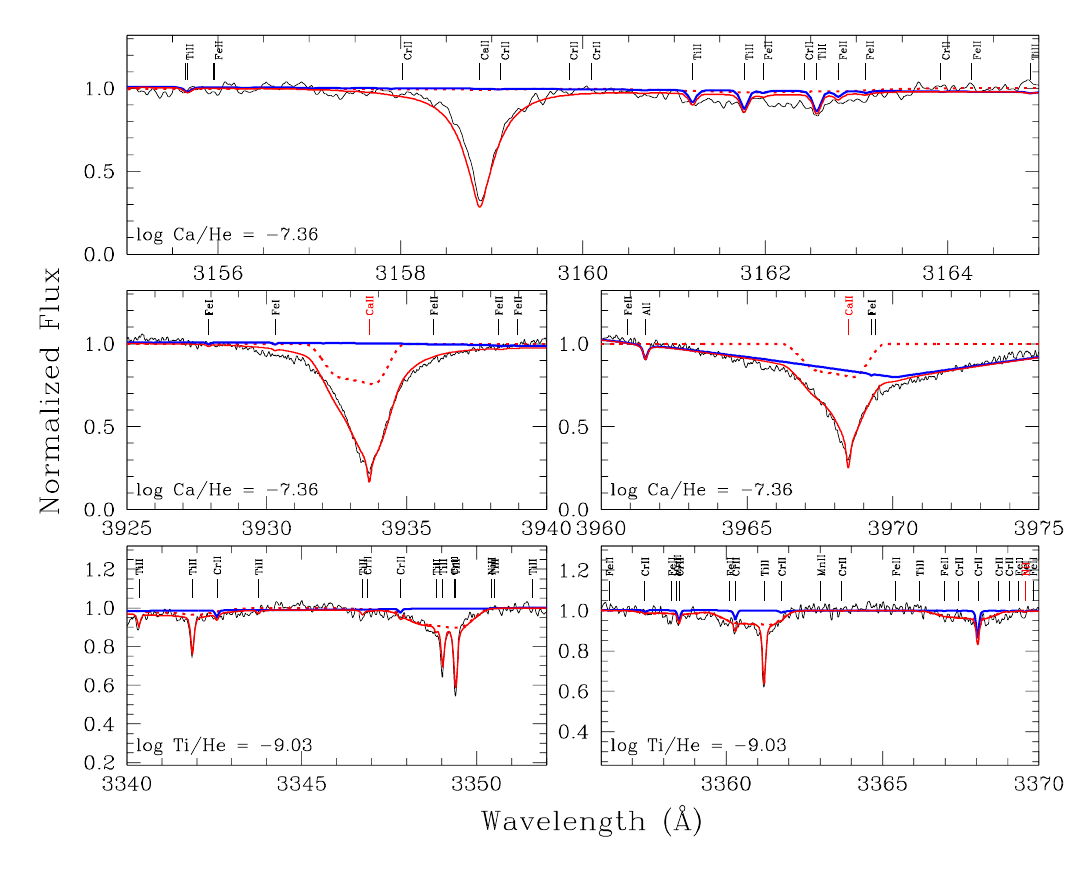}
\caption{Selected regions of the photospheric fit (red) to WD\,J0234-0406. In  dotted red is the expected contribution of the circumstellar gas disk for the velocity profile that matches best the Ti features (see Section~\ref{sec:disk-model}). The model photospheric + circumstellar gas spectrum without the fitted photospheric element is shown in blue. A red label is used for lines coming from the ground state.\label{fig8:fits02}}
\end{figure*}

\begin{figure*}[ht]
\includegraphics[trim={1cm 9.5cm 1.8cm 5.5cm},clip,width=\linewidth]{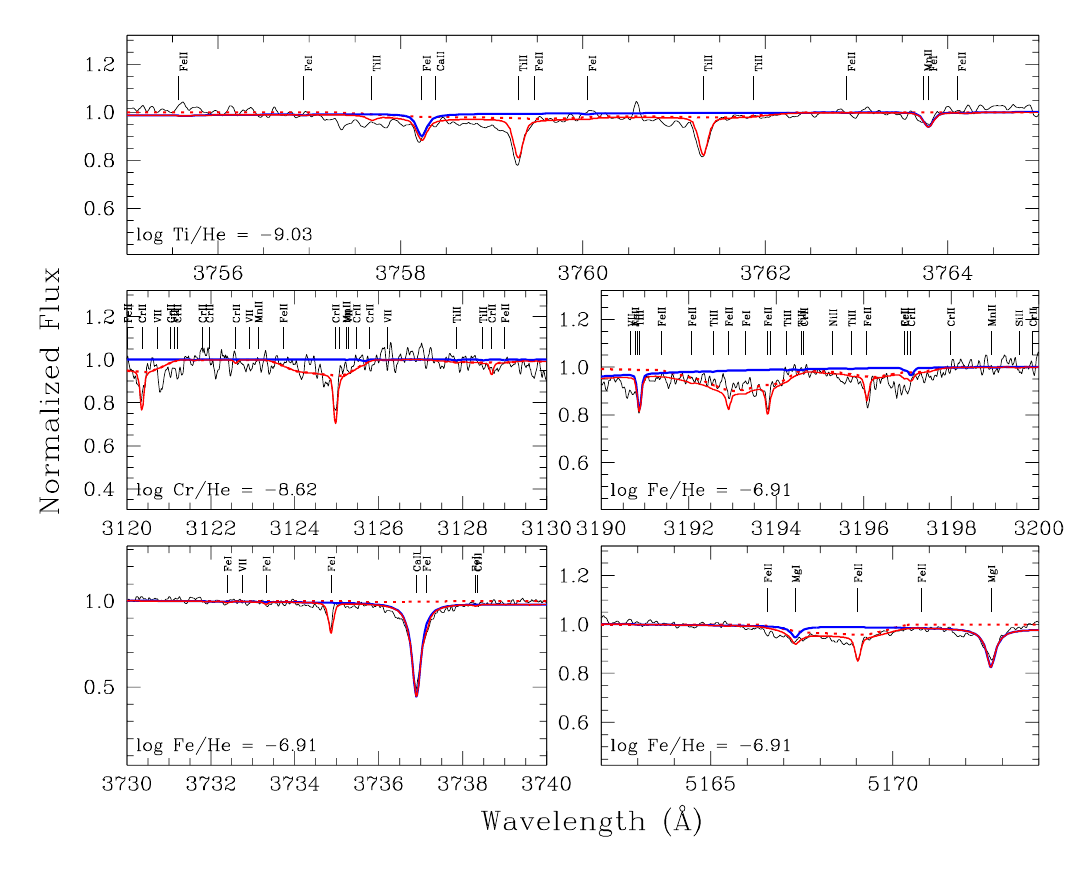}
\caption{Selected regions of the photospheric fit (red) to WD\,J0234-0406 and its circumstellar disk (dotted red). The model photospheric + circumstellar gas spectrum without the fitted photospheric element is shown in blue. A red label is used for lines coming from the ground state.\label{fig9:fits03}}
\end{figure*}

\begin{figure*}[ht]
\includegraphics[trim={1cm 9.9cm 1.8cm 5.5cm},clip,width=\linewidth]{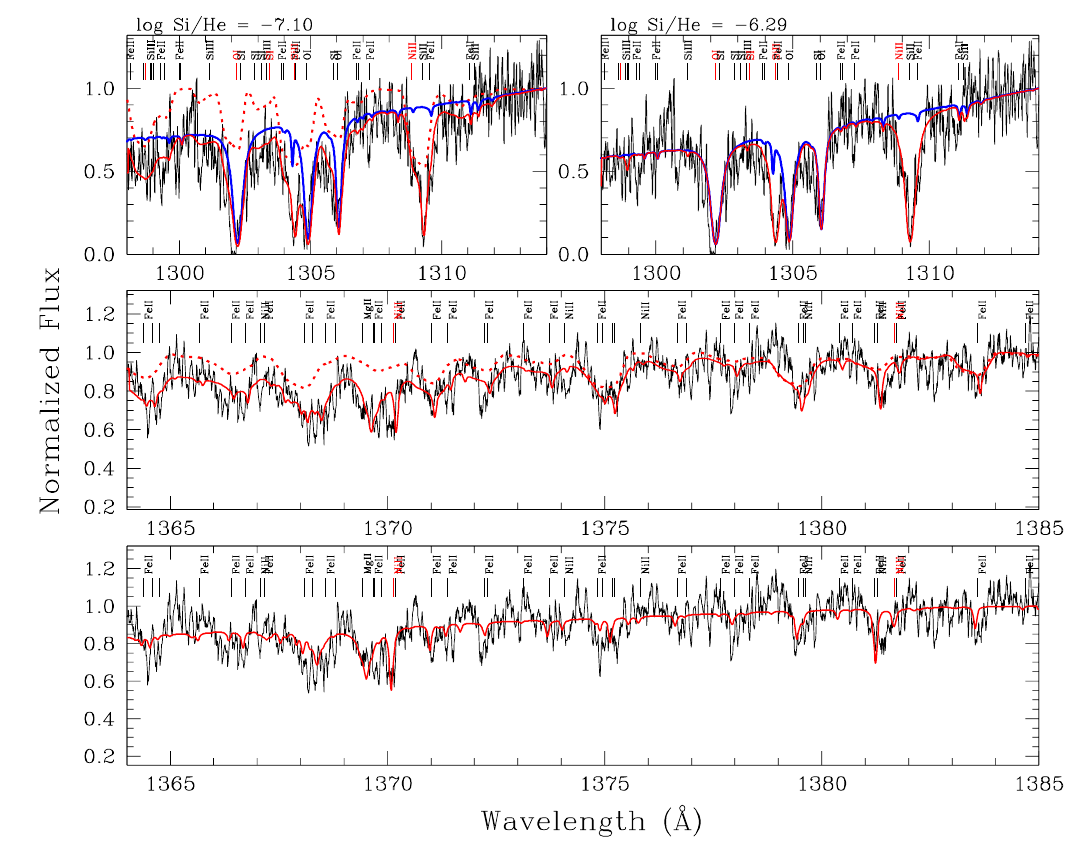}
\caption{Top left: Photospheric fit (red) of silicon lines in the UV, including the circumstellar disk (dotted red). Top right: Traditional photospheric fit (red) of silicon lines in the UV. The obtained Si abundance with (left) and without (right) including the disk are presented over the top two panels. The blue line represents the photosphere without silicon. Bottom: Selected region of the UV with (green) and without the contribution of the disk to the continuum. The rest wavelengths of the OI lines in the top two panels are 1302.17, 1304.86, and 1306.03 {\AA}.\label{fig10:disk-vs-no-disk}}
\end{figure*}

The optical and ultraviolet spectrum contains numerous photospheric lines from a variety of elements, see Figures~\ref{fig7:fits01} to \ref{fig10:disk-vs-no-disk} and Tables~\ref{tab5:optical} and \ref{tab6:uv}. The COS resolution is insufficient to resolve (separate) photospheric and circumstellar contributions to a given line. Photospheric abundances are given in Table~\ref{tab7:abn}. All optical abundances were derived using HIRES data only. Table~\ref{tab7:abn} also lists the abundances in the photosphere of WD\,1145+017 \citep{le_bourdais_revisiting_2024} and WD\,1425+540 as listed in Table 1 of \citet{xu_chemical_2017}. Based on its element abundances, the parent body accreted onto WD\,1425+540 was probably similar to a Kuiper Belt object of our solar system.

GF21 classify WD\,J0234-0406 as a DABZ, noting that although the star is He-rich the optical lines with the largest EW (by far) are Balmer lines. Since the white dwarf also displays lines from many heavy elements, the question is how to decide whether the secondary type should be DABZ or DAZB. \citet[see their Section 3.1]{doyle_new_2023} note that the paradigm established in \citet{sion_proposed_1983} and \citet{wesemael_atlas_1993} states that the spectral type is defined in order of the strongest optical spectral features, but no further definition is given as to what exactly this means. As noted by A. E. Doyle et al, if ``strong" refers to the depth of a line then the strongest line in a given spectrum can depend on linewidth and spectral resolution. Thus they suggest that equivalent width rather than line depth be used to define ``line strength", since EW is independent of line shape. The EW of the 5875.70 {\AA} \ion{He}{1} line in WD\,J0234-0406 is about the same as the EW of the \ion{Ca}{2} K-line. Therefore, the DABZ classification by GF21 is reasonable. Perhaps it is now time, 40 years on from the initial Sion et al paper, for the appearance of a new paper devoted to establishment of a white dwarf classification scheme (e.g. \citet{vincent_synthetic_2025}).

\begin{deluxetable*}{clllll}
\tablecaption{Photospheric Abundances in WD\,J0234-0406, WD\,1145+017 and WD\,1425+540.\label{tab7:abn}}
\tablehead{
\colhead{Element} & \colhead{WD\,J0234 (opt)} & \colhead{WD\,J0234 (UV)} & \colhead{WD\,1145 (opt)} & \colhead{WD\,1145 (UV)} & \colhead{WD\,1425} \\[-4pt]
 \colhead{} & \colhead{log n(Z)/n(He)} & \colhead{log n(Z)/n(He)} & \colhead{log n(Z)/n(He)} & \colhead{log n(Z)/n(He)} & \colhead{log n(Z)/n(He)}
 }
\startdata
Hydrogen & -1.97$\pm$0.20 & \nodata & -4.83$\pm$0.14 & \nodata & -4.20$\pm$0.10 \\
Beryllium & $<-11.5$ & \nodata & \nodata & \nodata & \nodata \\
Carbon & \nodata & -7.80$\pm$0.30 & \nodata & -7.39$\pm$0.29 & -7.29$\pm$0.17 \\
Nitrogen & \nodata & $<-7.0$ & \nodata & \nodata & -8.09$\pm$0.10 \\
Oxygen & -5.60$\pm$0.20 & -6.10$\pm$0.30 & -4.95$\pm$0.14 & -5.43$\pm$0.35 & -6.62$\pm$0.23 \\
Sodium & $<-8.0$ & \nodata & $<-6.0$ & \nodata & \nodata \\
Magnesium & -6.63$\pm$0.20 & \nodata & -5.75$\pm$0.28 & -5.98$\pm$0.26 & -8.16$\pm$0.20 \\
Aluminum & -7.60$\pm$0.20 & \nodata & -6.91$\pm$0.26 & -6.71$\pm$0.16 & \nodata \\
Silicon & -6.80$\pm$0.20 & -7.09$\pm$0.10 & -5.79$\pm$0.14 & -5.69$\pm$0.28 & -8.03$\pm$0.31 \\
Phosphorus & \nodata & $<-6.0$ & \nodata & -7.30$\pm$0.33 & \nodata \\
Sulfur & \nodata & $<-9.2$ & \nodata & -6.33$\pm$0.36 & -8.36$\pm$0.11 \\
Calcium & -7.36$\pm$0.20 & \nodata & -6.64$\pm$0.32 & \nodata & -9.26$\pm$0.10 \\
Titanium & -9.03$\pm$0.20 & \nodata & -8.39$\pm$0.22 & \nodata & \nodata \\
Vanadium & $<-10.0$ & \nodata & -9.01$\pm$0.40 & \nodata & \nodata \\
Chromium & -8.75$\pm$0.20 & \nodata & -7.62$\pm$0.28 & \nodata & \nodata \\
Manganese & -9.07$\pm$0.20 & \nodata & -8.34$\pm$0.32 & \nodata & \nodata \\
Iron & -6.91$\pm$0.14 & -7.33$\pm$0.20 & -5.39$\pm$0.32 & -5.46$\pm$0.31 & -8.15$\pm$0.14 \\
Nickel & \nodata & -8.41$\pm$0.20 & -6.61$\pm$0.37 & -6.79$\pm$0.32 & -9.67$\pm$0.20 \\
\enddata
\tablecomments{WD\,1145+017 abundances from \citet{le_bourdais_revisiting_2024}. WD\,1425+540 abundances from \citet{xu_chemical_2017}.}
\end{deluxetable*}

\subsection{Parent Body Composition and Mass}\label{sec4p5}

Elements heavier than helium settle out of white dwarf stellar photospheres in times much shorter than white dwarf cooling times (ages). Settling times are functions of the white dwarf effective temperature, surface gravity, and atmospheric composition, the latter usually assumed to be either pure H or pure He. As noted in Section~\ref{sec3p1}, the atmosphere of WD\,J0234-0406 is composed primarily of He, but with an unusually large fractional abundance of H. The presence of so much H substantially shortens the photospheric lifetime of heavy elements compared to what these lifetimes would be in a pure He atmosphere (see Table~\ref{tab8:mass}). 

\begin{deluxetable*}{lrrccccccc}\label{tab8:mass}
\tablecaption{Abundances and Parent Body Mass Compositions}
\tablehead{
Z & log($n$(Z)/$n$(He)) & log $\tau$(Z) & log $\tau$(Z) & \multicolumn{4}{c}{\% mass composition} \\
\cline{5-8}
& & (yrs) & (yrs) & Incr. Phase & St. State & & \\
& & Pure He & He+H mix & Any H/He & He+H mix & CI chondrites & Bulk Earth}
\startdata 
H & -1.97$\pm$0.20 & \nodata & \nodata &\nodata &\nodata & \nodata & \nodata\\
Be & $< -11.50$ & 6.391 &5.679 & \nodata & \nodata & $2.2\times 10^{-6}$ & $4.6\times 10^{-6}$ \\
C & -7.80$\pm$0.30 & 6.342 & 5.653 & $0.49^{+0.49}_{-0.25}$ & $0.36^{+0.36}_{-0.18}$ & 4.13 & 0.17 \\
N & $<-7.00$ & 6.283 & 5.626 & \nodata & \nodata & 0.25 & 0.0013 \\
O & -5.85$\pm$0.30 & 6.271 & 5.609 & $58.3 ^{+58.0}_{-29.1}$ & $47.5^{+47.2}_{-23.7}$ & 45.4 & 32.4 \\
Na & $< -8.00$ & 6.258 & $5.536 $& \nodata & \nodata & 0.51 & 0.19 \\
Mg & -6.63$\pm$0.20 & 6.276 & $5.534 $& $14.7 ^{+8.6}_{-5.4} $& $14.2 ^{+8.3}_{-5.3} $ & 9.5 & 15.8 \\
Al & -7.60$\pm$0.20 & 6.266 & $5.479 $& $1.7 ^{+1.0}_{-0.6} $& $1.9 ^{+1.1}_{-0.7} $ & 0.84 & 1.5 \\
Si & -6.95$\pm$0.20 & 6.280 & $5.466 $& $8.2 ^{+4.8}_{-3.0} $& $9.3 ^{+5.4}_{-3.4} $ & 10.8 & 17.1 \\
P & $< -6.00$ & 6.252 & $5.424 $& \nodata & \nodata & 0.10 & 0.07 \\
S & $< -9.20$ & 6.230 & $5.411 $& \nodata & \nodata & 5.4 & 0.46 \\
Ca & -7.36$\pm$0.20 & 6.099 & $5.372 $& $4.5 ^{+2.6}_{-1.7} $& $6.3 ^{+3.7}_{-2.3} $& 0.88 & 1.6 \\
Ti & -9.03$\pm$0.20 & 6.051 & $5.320 $& $0.12 ^{+0.07}_{-0.04} $& $0.18 ^{+0.11}_{-0.07} $& 0.05 & 0.08 \\
V & $< -10.00$ & 6.046 & $5.308 $& \nodata & \nodata & 0.005 & 0.009 \\
Cr & -8.75$\pm$0.20 & 6.062 & $5.313 $& $0.24 ^{+0.14}_{-0.09} $& $0.38 ^{+0.23}_{-0.14}$ & 0.26 & 0.42 \\
Mn & -9.07$\pm$0.20 & 6.061 & 5.296 & $0.12 ^{+0.07}_{-0.04}$ & $0.20 ^{+0.12}_{-0.07}$ & 0.19 & 0.14 \\
Fe & -7.12$\pm$0.21 & 6.079 & 5.291 & $10.9 ^{+6.8}_{-4.2} $ & $18.5 ^{+11.5}_{-7.1}$ & 18.6 & 28.8 \\
Ni & -8.41$\pm$0.20 & 6.101 & 5.274 & $0.59 ^{+0.34}_{-0.22}$ & $1.04 ^{+0.61}_{-0.38}$ & 1.10 & 1.7 \\
\hline
\multicolumn{4}{l}{Oxygen Budget} &0.46 $\pm$ 0.21&0.65 $\pm$ 0.29& & \\
\hline\enddata
\tablecomments{Abundances adopted for O, Si and Fe are the average of optical and UV derived values. $\tau$(Z) in the pure He column are the settling times in a $T_{\rm eff}=13,000~\rm K$, $\log g = 7.98$ WD atmosphere from the MWDD \citep{dufour_montreal_2017} in units of log(years). The log $\tau$(Z) listed in the He+H mix column are obtained from the average of the two settling times shown in Figure~\ref{fig15:koester_bedard}. The $\log \tau$(Z) Pure He column is included here to illustrate to what degree the presence of a substantial abundance of hydrogen in the convection zone reduces the settling times. The compositions by mass in the four columns on the right are the following: ``Incr. Phase" = directly observed abundance ratios, ``St. State" = abundance ratios corrected for differential element settling, ``CI chondrites" = meteoritic abundances from \citet{lodders_relative_2021}, ``bulk Earth" = compositions by mass from \citet{allegre_chemical_2001}. As defined in \citet{klein_chemical_2010}, an oxygen budget less than 1 implies there is excess oxygen not accounted for by rocky oxides (most likely originating from water in the parent body).} 
\end{deluxetable*}

\begin{figure}[h]
\centering
\includegraphics[trim={0 0 1.3cm 1cm},clip,width=\linewidth]{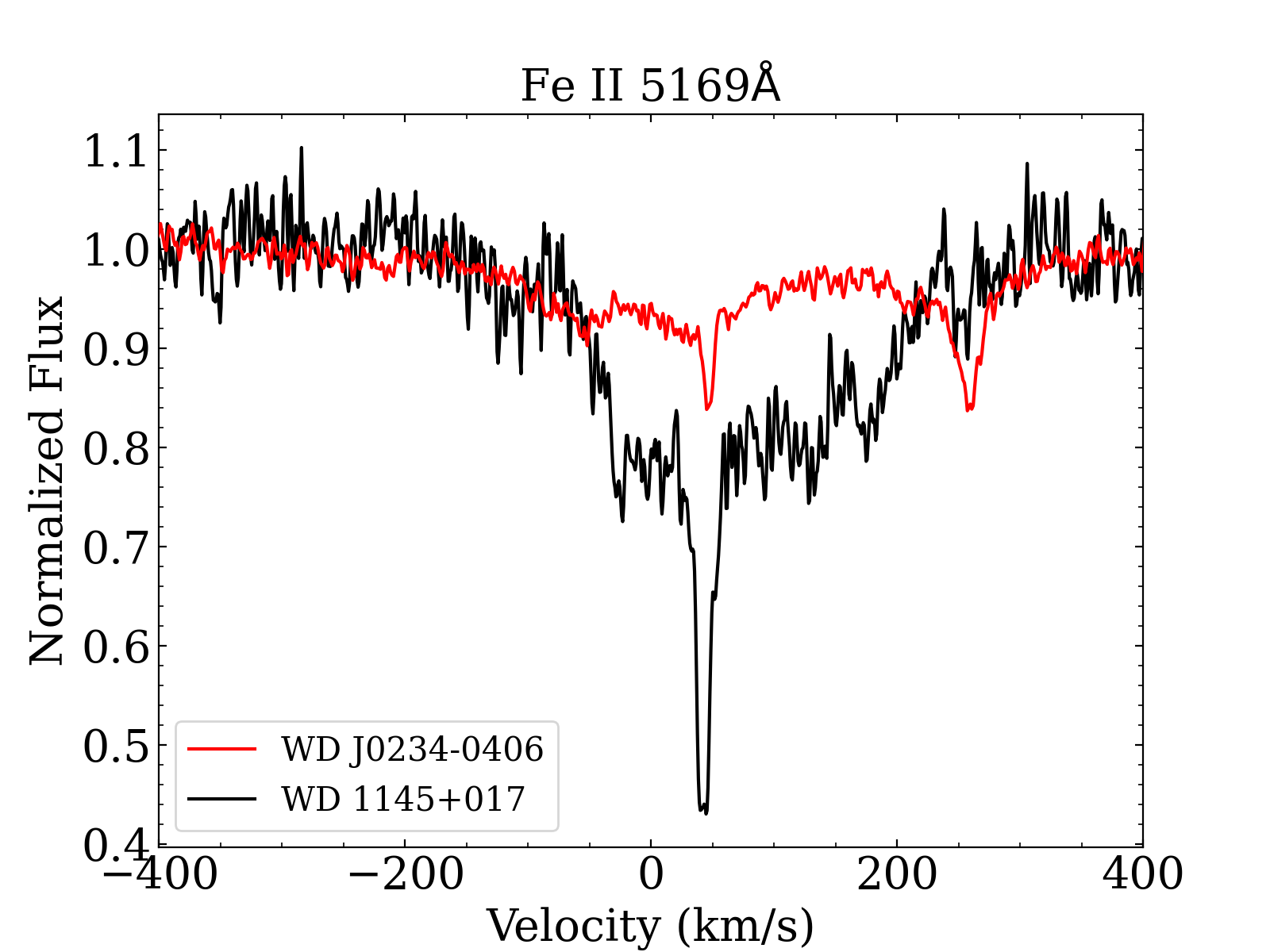}
\caption{\ion{Fe}{2} 5169 {\AA}.\label{fig13:wd1145}}
\end{figure}

\begin{figure}[h]
\centering
\includegraphics[width=\linewidth,trim={0 0.5cm 2cm 1.5cm},clip]{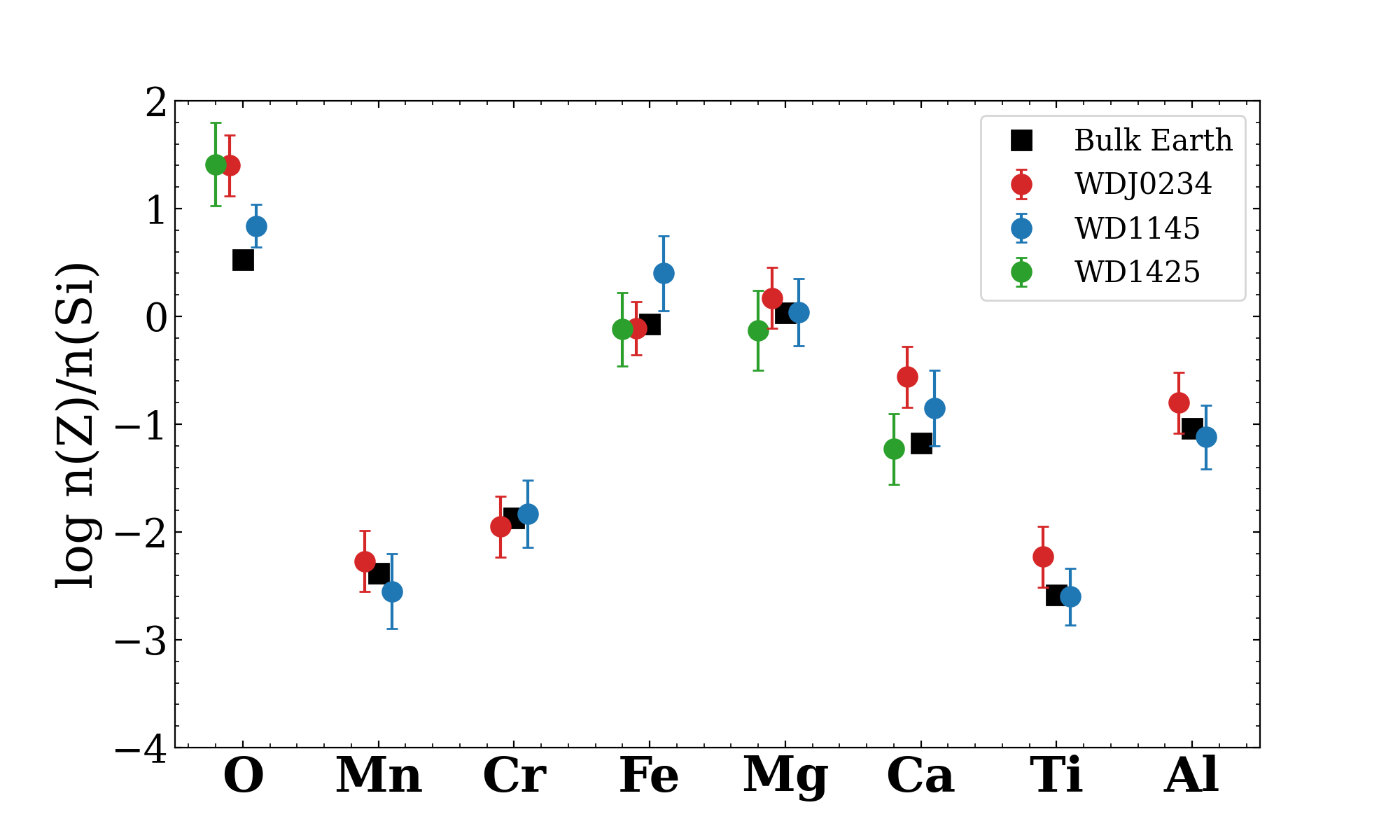}
\caption{Abundances comparison.\label{fig14:abn-comp}}
\end{figure}

As noted in the Introduction, WD\,J0234-0406 and WD\,1145+017 presently are the only white dwarfs to display broad circumstellar absorption lines. Figure~\ref{fig13:wd1145} illustrates how much deeper, typically, the broad absorption lines are in WD\,1145+017 compared to WD\,J0234-0406. The entries in Table~\ref{tab7:abn} show that heavy elements are substantially more abundant in the photosphere of WD\,1145+017; this is especially true for S, Fe, and Ni. For comparison, photopheric abundances for the Kuiper Belt like object accreted onto WD\,1425+540 are also listed. The Table~\ref{tab7:abn}  entries are displayed graphically in Figure~\ref{fig14:abn-comp}. 

\begin{figure}[h]
\centering
\includegraphics[width=\linewidth,trim={1.6cm 7.8cm 1.9cm 8.1cm},clip]{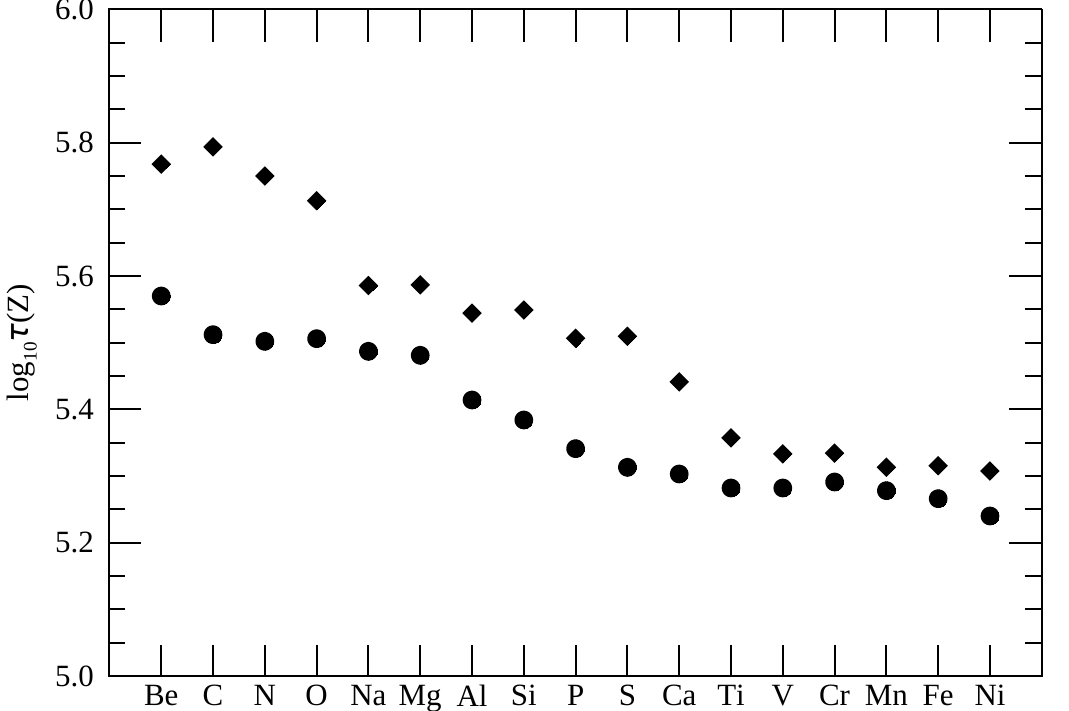}
\caption{Settling times in years for WD\,J0234$-$0406 as calculated using the Koester (diamonds) and Montreal (circles) white dwarf envelope codes. The differences in the settling times are mostly due to the different convective zone depths included in the two models.\label{fig15:koester_bedard}}
\end{figure}

WD\,J0234+0406 has a prominent circumstellar disk so either of two accretion situations are plausible. The so-called ``increasing phase" (Incr. Phase column in Table~\ref{tab8:mass}) pertains to the early stages of accretion of material from disk onto star; here there has been insufficient time for any accreted element to diffuse out of the convective zone and into the stellar interior. Based on the calculated settling times for the mixed H/He atmosphere (Figure~\ref{fig15:koester_bedard}), the increasing phase should last for a few times 10$^5$ years. The increasing phase is followed by the steady state (St. State in Table~\ref{tab8:mass}) where the rate of accretion of additional photospheric material from the disk is balanced by the rate at which heavy elements settle into the white dwarf interior and out of sight. 

The settling times listed in the fourth column of Table~\ref{tab8:mass} are an average of the two sets of calculated settling times displayed in Figure~\ref{fig15:koester_bedard}. The mass of the star in these models is 0.59 times the mass of the sun. For the Koester model, the log of the mass of the convective zone as a fraction of the stellar mass is -6.096, while for the Montreal model it is -6.345. These are roughly an order of magnitude smaller than the mass of the convective zone in a white dwarf with a pure He atmosphere and the same temperature and $\log g$ as WD\,J0234-0406.

We used two independent codes to calculate settling timescales for an atmospheric hydrogen abundance 1\% by number that of helium. One code is based on the method originally described in \citet{paquette_diffusion_1986}, with updates presented in \citet{fontaine_metal_2015} and \citet{heinonen_diffusion_2020}. This is the same framework used for the pure H and pure He timescales provided in the Montreal White Dwarf Database \citep{dufour_montreal_2017}. The other code is that described in \citet{koester_new_2020}. Both calculations assume no overshoot of the convection zone. We find that the presence of so much H substantially reduces the extent of the convection zone and thus shortens the photospheric lifetimes of heavy elements compared to the case of a pure He envelope. While the difference between settling times in the two models is about a factor of two, the agreement in relative settling times between two metals -- which is most important for the interpretation -- is significantly better.

The bottom row of Table~\ref{tab8:mass} gives the ``oxygen budget''. A detailed discussion of how the uncertainties in the oxygen budget are calculated may be found in Section 4.3 of \citet{klein_chemical_2010}. As noted in the last sentence of the caption to Table~\ref{tab8:mass}, the listed oxygen budgets of less than unity indicate that the parent body for the material currently seen in the photosphere contained a substantial quantity of water. Given the error bars on the oxygen budget, a dry parent body appears quite unlikely, but can't be ruled out with certainty. The unusually large percentage (1\%) of H in WD\,J0234+0406 contrasts with that seen in other white dwarfs with He dominated atmospheres; the H percentages listed in Table~\ref{tab7:abn} for WD\,1145+017 and WD\,1425+540 are much more typical. Because, by number, H is 4 orders of magnitude more abundant than O, only a tiny fraction of the H in the atmosphere of WD\,J0234-0406 could have been carried by the parent body responsible for the heavy elements now seen in the photosphere. Most of the H must have accompanied previously accreted objects or have been present when the white dwarf formed or, more likely, both. According to simulations of H transport in He-rich white dwarfs, primordial H likely contributes but cannot fully account for such a large photospheric abundance, thus requiring an external source \citep{rolland_spectral_2018,bedard_spectral_2023}. 

The current heavy element mass contained in the convective zone gives a lower limit to the mass of the parent body. This heavy element convective zone mass is about 10$^{22}$\,g which is rather smaller than the minimum parent body mass given for many He atmosphere white dwarfs (e.g., WD\,1145+017). But much of the difference can be accounted for by the relatively small mass of the WD\,J0234-0406 convective zone -- as noted two paragraphs above.

\citet{le_bourdais_revisiting_2024} conclude that in WD\,1145+017 ``both the disk and photosphere compositions align, to first order, with CI Chondrite." The same appears to be the case for WD\,J0234-0406; however the parent body of WD\,1145+017 apparently had a more prominent Ni-Fe core than that of WD\,J0234+0406. In addition, the volatile elements S and C in WD\,J0234-0406 are substantially less abundant than in chondritic material. Some models of Earth's core include these elements, although more commonly O and/or Si \citep{hirose_light_2021}.

\section{DISCUSSION}\label{sec5}

\subsection{Optical Thickness}\label{sec5p1}

At least some gaseous emission lines at white dwarf stars appear to be optically thick \citep[e.g.,][]{gansicke_gaseous_2006,horne_emission_1986}. Many of the circumstellar absorption lines at WD\,J0234-0406 are also likely to be optically thick formed by gas that covers $\leq$10$\%$ of the star because the circumstellar lines are never more than about 10\% deep, and lines with large EW are broader (Table~\ref{tab3:lines}), consistent with optical thickness. For WD\,1145+017, \citet{redfield_spectroscopic_2017} argue that the circumstellar lines are not very optically thick based on correlation of EW with $\log gf$ for \ion{Fe}{2} lines (their Figure~4). Inspection of Fe and Ti lines in our Table~\ref{tab3:lines} yields no obvious such correlation or correlation with lower state energy (E in Table~\ref{tab3:lines}).

\subsection{Radial Velocities}\label{sec5p2}

\citet{debes_detection_2012} report a white dwarf with a weak, narrow, circumstellar \ion{Ca}{2} K line blueshifted by $\sim$30 km s$^{-1}$ from the photospheric line; they interpret the velocity offset as due to gravitational redshift of the latter line. The circumstellar gas is traveling in a circular orbit with radius equal to 54 times the radius of a white dwarf.

For WD\,J0234-0406 with the temperature and gravity given in Section~\ref{sec3p1}, the evolutionary models of \citet{bedard_spectral_2020} predict a gravitational redshift of 28.6$\pm 0.4$ km s$^{-1}$. We measured the radial velocity of the photosphere of WD\,J0234-0406 using a few strong, uncontaminated lines of \ion{Ca}{2}, \ion{Ti}{2}, \ion{Fe}{1} and \ion{Mg}{1}. The velocities of all elements agree to within a km s$^{-1}$ or so of each other. We adopt 46$\pm$1 km s$^{-1}$ as the velocity of the photosphere, of which 28.6$\pm$0.4 km s$^{-1}$ is the gravitational redshift (Table~\ref{tab2:param}).

To measure the radial velocity of the circumstellar gas we used 12 lines of \ion{Fe}{1}, 8 lines of \ion{Ti}{2}, 4 lines each of \ion{Cr}{2} and \ion{Fe}{2}, and one line each of \ion{Ca}{1} and \ion{V}{1}; all lines had EW $\geq$90 m\AA. Because the circumstellar features are weak and broad the standard deviations of the measured velocities for the 4 ions with multiple measured lines are large; these standard deviations range between about 7 and 10 km s$^{-1}$. Of course the standard errors are much smaller than the standard deviations. In all cases, the velocities of the different ions agree with each other within the errors. We adopt 40$\pm$2 km s$^{-1}$ as the radial velocity of the circumstellar gas, while admitting the possibility that the radial velocities of different elements might differ from each other.

Thus the measured difference in radial velocity between the photospheric and circumstellar gas is about 6 km s$^{-1}$, which is much less than the gravitational redshift of 28.6 km s$^{-1}$. This small measured velocity difference is likely a consequence of more circumstellar gas along the line of sight to the star streaming away from Earth than toward it. This is consistent with the broad linewidths that imply very non-circular gas orbits \citep[e.g., Figure~4 in][]{fortin-archambault_modeling_2020}. Indeed, the circumstellar features seen at WD\,1145+017 can be redshifted with respect to the photospheric lines (see, e.g., Figure~1 in \citet{xu_evidence_2016} or Figure~5 in \citet{xu_shallow_2019}), notwithstanding the gravitational redshift of the latter lines.

\subsection{\ion{Si}{4}}\label{sec5p3}

\begin{figure*}[ht]
    \centering
    \includegraphics[width=\linewidth]{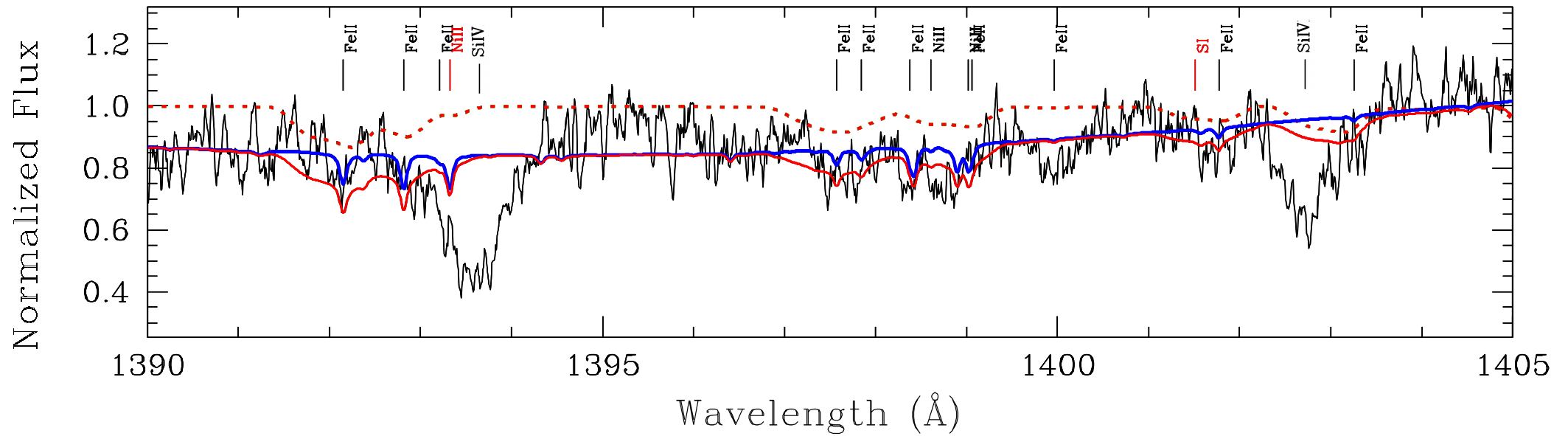}
    \caption{\ion{Si}{4} lines present in the spectrum of WD\,J0234-0406 at 1394 {\AA} and 1403 \AA. The line colors have the same meaning as in previous figures.}
   \label{fig16:si4}
\end{figure*}

Two transitions of \ion{Si}{4} are seen in absorption in the COS spectrum of WD\,J0234-0406 (Figure~\ref{fig16:si4} and Table~\ref{tab9:si4}). The existence of substantial numbers of \ion{Si}{1} ions in this high state of ionization at the 13,000 K photospheric temperature of this star is incompatible with our atmospheric models. In addition, the $\sim$700 m{\AA} full linewidth at half maximum intensity is far larger than the FWHM of photospheric lines and is characteristic of the circumstellar gas. Thus, the \ion{Si}{4} lines must be associated in some way with circumstellar material. According to the published literature and research currently underway (see Table~\ref{tab9:si4}), we are aware of a total of 11 white dwarfs with effective temperature $<$23,000 K that display these \ion{Si}{4} lines. Discussion of their significance and interpretation appears in Section 7 of \citet{gansicke_chemical_2012} and Section 3.2 of \citet{rogers_seven_2024a}.

As noted in Section~\ref{sec5p1}, optical circumstellar lines are never deeper than about 10\% of the stellar continuum, while in the UV the COS resolution is insufficient to cleanly separate circumstellar and photospheric line contributions. But the \ion{Si}{4} lines originate only in the circumstellar gas and, as may be seen in Figure~\ref{fig16:si4}, their depth is substantially greater than 10\%. Thus the \ion{Si}{4} ions must occlude a large fraction of the disk of the star and may be optically thick. The \ion{Si}{4} lines in PG 0843+516 \citet{gansicke_chemical_2012} and Gaia\,J0006+2858 \citet{rogers_seven_2024a} are also very deep indicating substantial optical depth and fractional coverage of the stellar disk. The close-in \ion{Si}{4} disk gas apparently covers more of the star than does more distant gas evident in other ions.

As noted in \citet{gansicke_chemical_2012}, the \ion{Si}{4} spectrum of WD\,1226+110 (= SDSS 1228+1040) is composed of two absorption components, one in the photosphere and a weaker, blueshifted, one carried by gaseous circumstellar material. The COS spectrum was obtained in 2010. The circumstellar feature is no longer present in COS spectra obtained in 2016, 2017 and 2018 (C. Manser 2024, personal communication). If, as seems plausible, the \ion{Si}{4} absorption is due to the hottest portion of a ring of circumstellar gas, then this hot material may have precessed away from the line of sight between Earth and WD\,1226+110.

\begin{deluxetable*}{lcccccccccc}
\tablecaption{\ion{Si}{4} Absorption Lines\label{tab9:si4}}

\tablehead{
Star & Atm. & $T_{\rm eff}$ & $\log g$ & Dist & RV Phot & 1394 \AA &  1394 \AA & 1403 \AA & 1403 \AA & Ref. \\
 & & (K) & & (pc) & (km/s) & RV (km/s) & EW (m\AA) & RV (km/s) & EW (m\AA) & 
 }
\startdata
Gaia\,J0006+2858 & H & 22,840 & 7.86 & 152 & 23.8 & -196 & \nodata & -196 & \nodata &  (1) \\
Gaia\,J0006+2858 & H & 22,840 & 7.86 & 152 & 23.8 & 20 & 213 (36) & 25.6 & 122 (45) & (1) \\
WD\,J0234-0406 & He & 13,000 & 7.98 & 91 & 46 & -7.5 & 538 & 29.9 & 425 & (2) \\
Gaia\,J0347+1624 & H & 18,850 & 7.84 & 141 & 16.0 & \nodata & \nodata & \nodata & \nodata & (1) \\
WD 0408-041 & H & 15,270 & 8.09 & 72 & 13.0 & -20.6 & 580 & -21.3 & 405 & (1, 1*) \\
Gaia\,J0510+2315 & H & 20,130 & 8.13 & 65 & 26.1 & \nodata & \nodata & \nodata & \nodata & (1) \\
Gaia\,J0611-6931 & H & 16,530 & 7.81 & 143 & 58.7 & \nodata & \nodata & \nodata & \nodata & (1) \\
Gaia\,J0644-0352 & He & 17,000 & 7.98 & 112 & 93.2 & \nodata & \nodata & \nodata & \nodata & (1) \\
Gaia\,J0649-7624 & H & 21,910 & 8.09 & 144 & 32.6 & 29.9 & 76 & 28.9 & 58 & (3) \\
WD 0843+516 & H & 23,097 & 8.17 & 139 & 1.2 & -13.5 & 826 & -13.5 & 600 & (4) \\
PG 1015+161 & H & 19,200 & 8.22 & 88 & 77.4 & \nodata & \nodata & \nodata & \nodata & (4) \\
SDSS J1043+0855 & H & 18,330 & 8.05 & 169 & 39.0 & \nodata & \nodata & \nodata & \nodata & (5) \\
WD\,1145+017 & He & 15,500 & 8.19 & 146 & 25 & -15 & 22 & -15 & 16 & (6, 7) \\
WD\,1226+110 & H & 20,900 & 8.15 & 129 & 41.3 & \nodata & \nodata & \nodata & \nodata & (4*) \\
WD\,1622+587 & He & 21,530 & 7.98 & 183 & -23.0 & -23.4 & 94 & -23.7 & 72 & (1)\\
GALEX J1931+0117 & H & 21,200 & 7.91 & 53 & 38.0 & 30.7 & 540 & 30.7 & 330 & (4) \\
SDSS J1957+3404 & H & 17,955 & 8.12 & 128 & -30.0 & -73.8 & 800 & -70.1 & 490 & (8) \\
Gaia\,J2100+2122 & H & 22,000 & 7.92 & 88 & 4.7 & \nodata & \nodata & \nodata & \nodata & (1) \\
\enddata
\tablecomments{References: (1) \citet{rogers_seven_2024a}; (1*) L. K. Rogers (2025, private communication); (2) this paper; (3) \citet{trierweiler_white_2025}; (4) \citet{gansicke_chemical_2012}; (4*) See text, Section~\ref{sec5p3}; (5) \citet{melis_does_2017}; (6) \citet{fortin-archambault_modeling_2020}; (7) \citet{le_bourdais_revisiting_2024}; (8) C. Melis (2025, private communication).} 
\end{deluxetable*}

Table~\ref{tab9:si4} reveals some regularities in the characteristics of the ensemble of detected \ion{Si}{4} lines. For a few of the hotter white dwarfs, the \ion{Si}{4} radial velocities are sometimes similar to those of the stellar photospheres thus suggesting possible origin in the photosphere. But usually \ion{Si}{4} is blue-shifted with respect to the photosphere, whatever the photospheric temperature. If the blue-shift is small then the \ion{Si}{4} may be formed in a hot gaseous layer with nearly circular orbit close to the star \citep{gansicke_chemical_2012}. Here, the difference in radial velocities could be due entirely to a somewhat larger gravitational redshift of the photosphere compared to the orbiting \ion{Si}{4} ions. But WD 0408-041 and SDSS J1957+3404 have \ion{Si}{4} blue-shifts of order 40 km s$^{-1}$. So, either the \ion{Si}{4} lines are somehow formed far enough from the stars so that there is negligible gravitational redshift or a kinematic component, i.e. some non-circular orbital motion, must be present. Such a situation is dramatically seen in Gaia\,J0006+2858 where \ion{Si}{4} absorption is blue-shifted from the photosphere by about 220 km s$^{-1}$ \citep[\ref{tab9:si4} and][]{rogers_seven_2024a}. This large shift combined with the modest width of this blue-shifted line (not smeared out in velocity as in circumstellar material at WD\,J0234-0406 and WD\,1145+017) suggests some kind of elliptical ring of material that passes close to Gaia\,J0006+2858.

\section{Conclusions}\label{sec6}

Previously, only one white dwarf (WD\,1145+017) with broad circumstellar absorption lines (100s of km/s wide) has appeared in the literature \citep{xu_evidence_2016}. This paper presents a second such example: WD\,J0234-0406. In order to account for the large linewidths the orbits of the circumstellar material must be substantially non-circular. In WD\,1145+017 the line profile changes dramatically with time \citep{fortin-archambault_modeling_2020, cauley_evidence_2018}, shifting from redshifted to blueshifted with respect to the photosphere. This has been modeled as due to precessing rings of high velocity gas \citep[][and references therein]{budaj_wd_2022}. Dramatic variations are also seen in the depths of dips in the broadband flux from WD\,1145+017. However, no obvious secular variations are seen in the spectrum of WD\,J0234-0406 -- neither photometric nor spectroscopic. Notwithstanding such differences, for WD\,J0234-0406 we utilized a gas disk model updated from that used by \citet{le_bourdais_revisiting_2024} for analysis of WD\,1145+017. The best fitting disk model indicates the presence of ions of at least a dozen different elements. None of the circumstellar absorption features (with the exception of two UV lines of \ion{Si}{4}) appears to be deeper than about 10\% of the continuum. Thus, the main portion of the circumstellar disk probably covers only about 10\% of the face of the white dwarf.

The elemental composition of the parent body responsible for the observed photospheric heavy elements, as well as the dusty/gaseous disk, appears to have been similar to that of CI chondrites, but with a lower abundance of volatile elements, specifically S and C. Of interest is whether the parent body carried a substantial quantity of water; to determine this we used a technique first employed by \citet{klein_chemical_2010} for He atmosphere white dwarfs. Our conclusion is that the parent body was likely quite ``wet", although given the size of the error bars on the abundances of the relevant elements, a dry parent body can't be ruled out with certainty. 

The photosphere of WD\,J0234+0406 is helium dominated, but with an unusually high fractional abundance of hydrogen (1\% by number). This represents so much H that even if the object currently being accreted was wet, its water content can account for only a minority of the hydrogen in the photosphere of the star. Therefore, the bulk of the observed H was obtained from parent bodies that accreted onto the star long ago, or is primordial, or both. 

\ion{Si}{4} lines are seen in the ultraviolet spectrum of WD\,J0234-0406; given the relatively low (13,000 K) temperature of the star these lines must be formed in the circumstellar material and not in the stellar photosphere. We have gathered existing \ion{Si}{4} spectra from other white dwarfs of which we are aware for comparison with WD\,J0234-0406. For some of the hotter stars, the \ion{Si}{4} lines might be formed in the stellar photosphere, but in most cases, more likely, from hot circumstellar material that orbits near the white dwarf. For a few stars, including WD\,J0234-0406, the \ion{Si}{4} lines appear to be formed in a layer of hot gas in an elliptical orbit that passes close to the stellar photosphere. In WD\,J0234-0406 and some other white dwarfs the \ion{Si}{4} absorption lines are very deep indicating that the absorbing gas must cover a major portion of the face of the star plus the Si line optical depths must be substantial. 

White dwarf stars now known to be accreting material onto their photospheres from their surrounding planet systems number more than 1700 \citep{williams_pewdd_2024}, a number that will soon greatly increase: ``The forthcoming DESI DR1 sample contains over 47\,000 white dwarf candidates, and will revolutionize statistical studies of white dwarf samples." \citep{manser_frequency_2024}. The number of candidate exoplanets around main sequence stars is of order 7000\footnote{\url{https://exoplanetarchive.ipac.caltech.edu/}}. Typically, the white dwarfs sample stars more massive and older than the main-sequence stars along with planetary material with larger orbital semi-major axes.\\ 

\begin{acknowledgments}
We thank the referee for comments/suggestions that helped to improve the paper. We are grateful to Boris G\"{a}nsicke, Snehalata Sahu, Chris Manser, and Laura Rogers for kindly supplying us with unpublished data or calculations and we thank Edward Young and Jay Farihi for helpful advice. C.M.\ and B.Z.\ acknowledge support from NSF grants SPG-1826583 and SPG-1826550. B.K. acknowledges support from NASA grant HST-GO-16204.005-A to UCLA.

This paper is based on observations made with the NASA/ESA Hubble Space Telescope, obtained at the Space Telescope Science Institute, which is operated by the Association of Universities for Research in Astronomy, Inc., under NASA contract NAS 5-26555.
This work was supported by the associated grant HST-GO-15461.001-A. 
Research at Lick Observatory is partially supported by a generous gift from Google.
This paper includes data gathered with the 6.5 m Magellan Telescopes located at Las Campanas Observatory, Chile.

Some of the data presented herein were obtained at the W.M. Keck Observatory, which is operated as a scientific partnership among the California Institute of Technology, the University of California, and NASA. The Observatory was made possible by the generous financial support of the W. M. Keck Foundation. We recognize and acknowledge the very significant cultural role and reverence that the summit of Maunakea has always had within the indigenous Hawaiian community. S. Xu is supported by the international Gemini Observatory, a program of NSF NOIRLab, which is managed by the Association of Universities for Research in Astronomy (AURA) under a cooperative agreement with the US National Science Foundation, on behalf of the Gemini partnership of Argentina, Brazil, Canada, Chile, the Republic of Korea, and the United States of America.
\end{acknowledgments}

\bibliography{references}
\bibliographystyle{aasjournal}
\end{document}